\documentclass[11pt,a4paper]{article}
\usepackage[utf8]{inputenc}
\usepackage{a4wide}
\usepackage{amsmath,amssymb}
\usepackage{amsfonts}
\usepackage{cite}
\usepackage{color}
\usepackage{adjustbox}
\usepackage{multirow}
\usepackage{pdflscape}
\usepackage{textcomp}
\usepackage{float}
\usepackage{gensymb}
\usepackage{graphicx}
\usepackage{diagbox}
\usepackage{pifont}
\usepackage{bigstrut}
\usepackage[small,bf]{caption}
\usepackage{tabulary}
\usepackage{booktabs}
\usepackage{stackengine}
\usepackage{mathtools}
\usepackage{enumerate}
\usepackage{makecell}
\usepackage{listings}
\usepackage{bbm}
\usepackage[unicode]{hyperref} 

\definecolor{greenLinks}{rgb}{0, 0.6, 0} 
\definecolor{blueLinks}{rgb}{0, 0, 0.6}
\definecolor{redLinks}{rgb}{0.6, 0, 0}
\definecolor{eprintLinks}{rgb}{0.4, 0.4, 0.4}
\definecolor{journalLinks}{rgb}{0.6, 0, 0}
\begin{document}

\pagenumbering{Alph}
\begin{titlepage}

\vspace*{15mm}

\begin{center}
{ \bf\LARGE {An  $SU(5)$ GUT with non-holomorphic modular $S'_4$ \\[2mm] symmetry: fermion masses, mixing and leptogenesis}}\\[8mm]
Xianshuo Zhang$^{\,a,}$\footnote{E-mail: \texttt{zxianshuo@yeah.net}},
Yakefu Reyimuaji$^{\,a,}$\footnote{E-mail: \texttt{yreyi@hotmail.com}} \\
\vspace{8mm}
$^{a}$\,{\it School of Physical Science and Technology, Xinjiang University, Urumqi, Xinjiang 830017, China} \\
\vspace{2mm}
\end{center}
\setcounter{footnote}{0} 

\vspace{8mm}

\begin{abstract}
\noindent  
A non-supersymmetric $SU(5)$ grand unified model is constructed based on the non-holomorphic modular $S'_4$ flavor symmetry, supplemented by a global Peccei--Quinn $U(1)_{PQ}$ symmetry that constrains the Yukawa couplings. Constructed from polyharmonic Maa{\ss} forms, these couplings allow the model to reproduce the observed charged-fermion masses and mixing as well as the neutrino oscillation parameters for both normal ordering (NO) and inverted ordering (IO), with all Pull values satisfying $|\mathcal{P}| < 0.8$. The model further yields specific predictions for yet-unmeasured parameters, including the absolute neutrino masses, the Dirac CP phase, the Majorana phases, and the effective Majorana mass $m_{\beta\beta}$. We then study thermal leptogenesis by solving the density matrix equations. For a given benchmark point, the final baryon asymmetry depends solely on the single free parameter $b$ that sets the right-handed neutrino mass scale, and an appropriate choice of this parameter reproduces the observed value $\eta_B \simeq 6.12 \times 10^{-10}$. Finally, the correlation between this scale parameter and specific low-energy observables provides a potential bridge between the fermion flavor structure and early-Universe cosmology.

\end{abstract}


\end{titlepage}
\pagenumbering{arabic}



\section{Introduction}
\label{sec:intro}
The Standard Model (SM) has passed stringent experimental tests and is a highly successful theory of the strong and electroweak interactions. Nevertheless, it does not explain the origin of fermion masses and flavor mixing. It offers no natural explanation for the observed hierarchy of fermion masses, nor can it account for the distinct mixing patterns in the quark and lepton sectors. These observations point to an underlying flavor structure. In addition, neutrino oscillation experiments have established that neutrinos have tiny masses, and cosmological observations show that the Universe exhibits a matter-antimatter asymmetry. Explaining these puzzles, and understanding whether they are interconnected, is a central motivation for physics beyond the SM (BSM).

Grand unified theories (GUTs) provide an attractive framework for addressing some of these issues~\cite{Pati1974,GeorgiGlashow1974,Georgih1974,Georgi1975StateOfArt,FritzschMinkowski1975,Langacker2012}. By embedding the SM gauge group into a larger one, GUTs unify quarks and leptons into common gauge multiplets and can establish relations among their Yukawa couplings. In particular, $SU(5)$ offers a simple realization of this idea, with the SM fermions embedded in the $\overline{\mathbf{5}}$ and $\mathbf{10}$ representations~\cite{GeorgiGlashow1974,Georgih1974,Langacker2012}.

Although the minimal $SU(5)$ model does not include right-handed neutrinos, they can be introduced as additional gauge singlets $N_i$, which naturally explain the tiny neutrino masses through the seesaw mechanism~\cite{MINKOWSKI1977421,GellMannRamondSlansky1979,Yanagida1979,MohapatraSenjanovic1980,Mohapatra:2006gs}. The out-of-equilibrium and CP-violating decays of these heavy singlets can then violate lepton number and generate a lepton asymmetry. It is subsequently converted into the observed baryon asymmetry by electroweak sphaleron processes, providing a unified framework for the origin of the cosmic baryon asymmetry~\cite{Fukugita:1986hr}. Successful leptogenesis has indeed been realized in numerous GUT models~\cite{Pati:2002pe,Abada:2008gs,Kannike:2011fx,Fong:2014gea,Gehrlein:2015dxa,Ishihara:2015uua,Bjorkeroth:2016lzs,King:2024gsd}.

However, to correct the mass relations between down-type quarks and charged leptons, $SU(5)$ models generally require the introduction of the $\mathbf{45}_H$ Higgs representation. This often induces dangerous couplings that lead to rapid proton decay. To naturally forbid these unwanted interactions and ensure proton stability, a global $U(1)_{PQ}$ symmetry can be introduced~\cite{FileviezPerez:2019ssf}.

Flavor symmetries have also been widely employed to explain the observed patterns of fermion masses and mixing~\cite{FroggattNielsen1979,Altarelli:2010gt,Feruglio:2015jfa}. In conventional flavor models, the flavor symmetry is typically broken spontaneously by the vacuum expectation values (VEVs) of additional scalar fields, commonly called flavons. However, the required vacuum alignments of these fields often make the construction of realistic models complicated. Modular flavor symmetry offers an alternative. The Yukawa couplings are described by modular forms, and the VEV of the modulus parameter $\tau$ determines the fermion flavor structure~\cite{Feruglio:2017spp,Feruglio:2019ybq,Ding:2023htn}. This approach significantly reduces the reliance on extra flavon fields and lowers the number of free parameters.

The connection between modular flavor symmetry and leptogenesis has also attracted considerable attention~\cite{Asaka:2019vev,Okada:2021qdf,Kashav:2021zir,Qu:2021jdy,Kang:2022psa,Ding:2022bzs,Mishra:2022egy,Marciano:2024nwm,Singh:2024imk,Ding:2024euc,Abhishek:2025ety,Priya:2025wdm,Kumar:2025bfe,Zhang:2026kyy,Priya:2026ehe,Li:2026ory}. In such models, the modular forms determine the structure of the neutrino Yukawa couplings, and the modulus $\tau$ can simultaneously influence both the low-energy neutrino masses and mixing and the generation of the CP asymmetry responsible for leptogenesis.

GUTs and modular flavor symmetries have traditionally been formulated within the framework of supersymmetry (SUSY). Although SUSY remains theoretically appealing, no conclusive experimental evidence for supersymmetric particles has so far been found, which has motivated the exploration of models that do not rely on it. Non-supersymmetric (non-SUSY) GUTs provide an alternative framework in which fermion masses and mixing, gauge coupling unification (GCU), proton decay, and the origin of the baryon asymmetry can be studied within a common setting. In particular, successful leptogenesis has been achieved in various non-SUSY GUT models based on $SU(5)$ and $SO(10)$~\cite{Kannike:2011fx,Buccella:2012kc,Mummidi:2021anm,Fu:2022lrn,Fonseca:2023per}.

Modular symmetries have also been explored in non-SUSY scenarios, although modular flavor symmetry was originally introduced in the supersymmetric context, where the Yukawa couplings are described by holomorphic modular forms~\cite{Feruglio:2017spp}. Recent studies have extended modular forms to more general non-holomorphic forms, in which the Yukawa couplings are constructed from polyharmonic Maa{\ss} forms~\cite{Qu:2024rns,Qu:2025ddz}. This generalization allows modular forms with arbitrary integer weights and provides a flexible framework for constructing modular-invariant flavor models. Several studies have demonstrated the viability of non-holomorphic modular symmetries in explaining fermion masses and mixing~\cite{Qu:2024rns, Kumar:2024uxn, Nomura:2024atp, Nomura:2024nwh, Kobayashi:2025hnc, Loualidi:2025tgw, Kumar:2025bfe, Nomura:2025ovm, Nomura:2025raf, Zhang:2025dsa, Priya:2025wdm, Kumar:2025nut, Nanda:2025lem, Okada:2025nap, Gao:2025jlw,Li:2025kcr}. However, studies applying modular symmetry to non-SUSY GUTs remain relatively limited. Ref.~\cite{Loualidi:2025tgw} investigated $SU(5)$ GUTs based on the $A_4$ modular group but did not explore leptogenesis.

Motivated by these developments, we construct a non-SUSY grand unified model based on the non-holomorphic $S'_4$ modular flavor symmetry. This framework aims to provide a unified description of quark and lepton masses and mixing while connecting the flavor structure to the origin of the baryon asymmetry through thermal leptogenesis. Here, the modular symmetry constrains the Yukawa structure of the fermion sector, whereas the right-handed neutrino sector provides the necessary ingredients for thermal leptogenesis. We explore the interplay between fermion flavor mixing and the cosmic baryon asymmetry within this framework.

The remainder of this paper is organized as follows. In Sec.~\ref{sec:model}, we introduce the model, discuss its particle content and symmetry assignments, and present the fermion mass matrices. In Sec.~\ref{sec:fit}, we analyze fermion masses and mixing through a numerical fit. In Sec.~\ref{sec:tlep}, we investigate thermal leptogenesis. We summarize our results in Sec.~\ref{sec:conclusions}. Finally, Appendix~\ref{sec:gutpq} discusses the phenomenological consequences of the model, including gauge coupling unification, proton decay, and the axion mass and its implications, which follow from the additional $U(1)_{PQ}$ symmetry. Appendix~\ref{app:modular} reviews the fundamentals of modular symmetry.

\section{The model}
\label{sec:model}
In this section, we construct a non-SUSY, renormalizable $SU(5)$ grand unified model. The three generations of SM matter fields are embedded in the $\bar{\mathbf{5}}$ and $\mathbf{10}$ representations of $SU(5)$, denoted by $\bar{F}$ and $T$, respectively. The model also includes three $SU(5)$ singlet right-handed neutrinos $N_i$, which explain the small neutrino masses through the type-I seesaw mechanism. The scalar sector contains $H_5$, $H_{24}$, and $H_{45}$, which belong to the $\mathbf{5}$, $\mathbf{24}$, and $\mathbf{45}$ representations, respectively. $H_{24}$ spontaneously breaks the $SU(5)$ gauge symmetry, while $H_{45}$ provides the Yukawa structure needed to correct the mass relations between down-type quarks and charged leptons through the Georgi--Jarlskog factor of $-3$~\cite{Georgi:1979df}.

However, the $SU(5)$ gauge symmetry alone is not sufficient to fully constrain the form of the Yukawa interactions. In particular, couplings of the $TTH_{45}$ type may introduce additional colored-scalar interactions and affect proton stability. Inspired by Ref.~\cite{FileviezPerez:2019ssf}, we introduce a global Peccei--Quinn symmetry $U(1)_{\mathrm{PQ}}$. Although this symmetry also has implications for the strong CP problem and axion dark matter, we focus here on its role as a selection rule that constrains the flavor structure and ensures proton stability. Its phenomenological implications are discussed further in Appendix~\ref{sec:gutpq}. To make the up-type quark Yukawa interactions compatible with the above PQ selection rules, we introduce an additional scalar $H'_5$, which also belongs to the fundamental representation $\mathbf{5}$ of $SU(5)$. General PQ charge assignments can be found in Ref.~\cite{FileviezPerez:2019ssf}.

To describe the flavor structure among different generations of matter fields, we introduce a modular symmetry $S'_4$ on top of $SU(5)\times U(1)_{\mathrm{PQ}}$. The $SU(5)$ gauge symmetry unifies the different quark and lepton components within the same generation. The $S'_4$ symmetry, through the representation assignments of the matter fields and the constraints imposed by modular forms on the Yukawa couplings, determines the mass matrix structure of the different generations. The general form of the quark and lepton mass Lagrangian allowed by the $SU(5)\times U(1)_{\mathrm{PQ}}\times S'_4$ symmetry is
\begin{align}
\mathcal{L}={}&NNf_1(Y)+\bar{F}NH_5f_2(Y)+\bar{F}NH_5f_3(Y)\notag\\
&+\bar{F}TH^\ast_5f_4(Y)+\bar{F}TH^\ast_{45}f_5(Y)+TTH'_{5}f_6(Y)+\mathrm{H.c.},
\label{gl}
\end{align}
where $f_1(Y)$, $f_2(Y)$, $\cdots$, $f_6(Y)$ denote functions of the modular forms.
\begin{table}[ht]
  \begin{center}
    \renewcommand{\arraystretch}{1.35}
    \setlength{\tabcolsep}{8pt}
    \begin{tabular}{|c||c|c|c|c|c||c|c|c|}
      \toprule
      \text{Fields}
      & $T_1$& $T_2$& $T_3$& $\bar{F}_{1,2,3}$ & $N_{1,2,3}$& $H_5$& $H_{45}$& $H'_5$\\
      \hline
      $SU(5)$& $\mathbf{10}$& $\mathbf{10}$& $\mathbf{10}$& $\overline{\mathbf{5}}$ & $\mathbf{1}$ & $\mathbf{5}$& $\mathbf{45}$ & $\mathbf{5}$ \\
      \hline
      $S'_4$& $\hat{\mathbf{1}}$ & $\hat{\mathbf{1}}'$& $\mathbf{1'}$& $\mathbf{3}'$& $\mathbf{3}$& $\mathbf{1}$& $\mathbf{1}$& $\mathbf{1}$ \\
      \hline
      $U(1)_{PQ}$& $2$ & $2$ & $2$ & $-1$& $0$& $1$& $1$& $-4$ \\
      \hline
      $k$& $-1$& $-7$& $-8$& $4$& $0$ & $0$& $0$ & $12$ \\
      \bottomrule
    \end{tabular}
    \caption{Field assignments under the $SU(5)$ gauge symmetry, the modular flavor symmetry $S'_4$, and the Peccei--Quinn symmetry $U(1)_{PQ}$. The fields $T_i$, $\bar{F}_i$, $N_i$, $H_5$, $H_{45}$, and $H'_5$ transform under $SU(5)$ and $S'_4$ as indicated, and the last row lists their modular weights $k$.}
    \label{tab:field_assignment_PQ}
  \end{center}
\end{table}

In the concrete model constructed here, $\bar{F}=(\bar{F}_1,\bar{F}_2,\bar{F}_3)$ is assigned to the $\mathbf{3}'$ representation of $S'_4$, while $T=(T_1,T_2,T_3)$ decomposes as the direct sum $\hat{\mathbf{1}}\oplus\hat{\mathbf{1}}'\oplus\mathbf{1'}$. The right-handed neutrinos $N=(N_1,N_2,N_3)$ are assigned to the triplet $\mathbf{3}$ of $S'_4$. In the scalar sector, all the fields $H$ are trivial singlets $\mathbf{1}$ of $S'_4$. The assignments of these fields, together with their $U(1)_{\mathrm{PQ}}$ charges and modular weights, are summarized in Table~\ref{tab:field_assignment_PQ}. This choice of PQ charges forbids the $TTH_{45}$ term. In addition, $H_{24}$ is an $S'_4$ singlet carrying a PQ charge of $5$, and its modular weight can be chosen to be any even integer~\footnote{The VEV of $H_{24}$ simultaneously breaks the $SU(5)\times U(1)_{\mathrm{PQ}}$ symmetry, while the modular symmetry is broken by the VEV of the modulus field $\tau$.}. Since the modular weights of $H_5$, $H_{45}$, and $H'_5$ are also even, the total modular weights of the scalar interaction terms built from these fields and $H_{24}$ can likewise be chosen to be even. For the modular group $S'_4$, modular forms of even weight include singlet representations $Y_{\mathbf{1}}^{(\mathrm{even})}$. These scalar interactions are therefore not forbidden by modular symmetry.

In this model, the mass matrices of the charged leptons and down-type quarks are determined by the two types of Yukawa interactions in Eq.~\eqref{gl}, namely $\bar{F}TH_5^\ast f_4(Y)$ and $\bar{F}TH_{45}^\ast f_5(Y)$. With the chosen field assignments, the corresponding Lagrangian reads
\begin{equation}
    \begin{aligned}
         -\mathcal{L}_{d}
             ={}&\alpha_1(\bar{F}T_1)_{\hat{\mathbf{3}}'}Y^{(-3)}_{\hat{\mathbf{3}}}(\tau)H_5^\ast+ \alpha_2(\bar{F}T_2)_{\hat{\mathbf{3}}}Y^{(3)}_{\hat{\mathbf{3}}'}(\tau)H_5^\ast+\alpha_3(\bar{F}T_3)_{\mathbf{3}}Y^{(4)}_{\mathbf{3}}(\tau)H_5^\ast\\
             &+\alpha'_1(\bar{F}T_1)_{\hat{\mathbf{3}}'}Y^{(-3)}_{\hat{\mathbf{3}}}(\tau)H_{45}^\ast +\alpha'_2(\bar{F}T_2)_{\hat{\mathbf{3}}}Y^{(3)}_{\hat{\mathbf{3}}'}(\tau)H_{45}^\ast+\alpha'_3(\bar{F}T_3)_{\mathbf{3}}Y^{(4)}_{\mathbf{3}}(\tau)H_{45}^\ast +\mathrm{H.c.},
    \end{aligned}
\end{equation}
where $\alpha_i$ and $\alpha'_i$ are coupling constants and $Y^{(k)}_{\mathbf{r}}(\tau)$ denotes a modular form of weight $k$ transforming in the representation $\mathbf{r}$ of $S'_4$. The resulting Yukawa coupling matrices are
\begin{equation}
    y^{d}_{\bar{5}}=
            \begin{bmatrix}
                \alpha_1Y^{(-3)}_{\hat{\mathbf{3}},1}&\alpha_1Y^{(-3)}_{\hat{\mathbf{3}},3}&\alpha_1Y^{(-3)}_{\hat{\mathbf{3}},2}\\
                \alpha_2Y^{(3)}_{\hat{\mathbf{3}}',1}&\alpha_2Y^{(3)}_{\hat{\mathbf{3}}',3}&\alpha_2Y^{(3)}_{\hat{\mathbf{3}}',2}\\
                \alpha_3Y^{(4)}_{\mathbf{3},1}&\alpha_3Y^{(4)}_{\mathbf{3},3}&\alpha_3Y^{(4)}_{\mathbf{3},2}
            \end{bmatrix},
\end{equation}
and
\begin{equation}
    y^{d}_{\bar{45}}=
        \begin{bmatrix}
            \alpha'_1Y^{(-3)}_{\hat{\mathbf{3}},1}&\alpha'_1Y^{(-3)}_{\hat{\mathbf{3}},3}&\alpha'_1Y^{(-3)}_{\hat{\mathbf{3}},2}\\
            \alpha'_2Y^{(3)}_{\hat{\mathbf{3}}',1}&\alpha'_2Y^{(3)}_{\hat{\mathbf{3}}',3}&\alpha'_2Y^{(3)}_{\hat{\mathbf{3}}',2}\\
            \alpha'_3Y^{(4)}_{\mathbf{3},1}&\alpha'_3Y^{(4)}_{\mathbf{3},3}&\alpha'_3Y^{(4)}_{\mathbf{3},2}
        \end{bmatrix}.
\end{equation}
These two matrices describe the Yukawa structures associated with $H_5^\ast$ and $H_{45}^\ast$, respectively.

For the up-type quark sector, the masses arise from $TTH'_5f_6(Y)$. The explicit Lagrangian reads
\begin{equation}
    -\mathcal{L}_{u}=\beta_1(T_1T_2+T_2T_1)_{\mathbf{1}}Y^{(-4)}_{\mathbf{1}}(\tau)H'_5+\beta_2(T_2T_3+T_3T_2)_{\hat{\mathbf{1}}}Y^{(3)}_{\hat{\mathbf{1}}'}(\tau)H'_5+\beta_3(T_3T_3)_{\mathbf{1}}Y^{(4)}_{\mathbf{1}}(\tau)H'_5+\mathrm{H.c.},
\end{equation}
and the corresponding Yukawa coupling matrix is
\begin{equation}
    y^{u}_{5'}=
        \begin{bmatrix}
            0&\beta_1Y^{(-4)}_{\mathbf{1}}&0\\
            \beta_1Y^{(-4)}_{\mathbf{1}}&0&\beta_2Y^{(3)}_{\hat{\mathbf{1}}'}\\
            0&\beta_2Y^{(3)}_{\hat{\mathbf{1}}'}&\beta_3Y^{(4)}_{\mathbf{1}}
        \end{bmatrix}.
\end{equation}

The Higgs sector contains $H_5$, $H_{45}$, $H'_5$, and $H_{24}$. The first three $SU(5)$ multiplets contain the electroweak Higgs doublets
\begin{equation}
H_1\subset H_5,\qquad H_2\subset H_{45},\qquad H_3\subset H'_{5},
\end{equation}
with $H_1,\ H_2,\ H_3 \sim (1,2,+1/2)$. These doublets mix to form the Higgs doublets of the low-energy effective theory. Their VEVs $v_1$, $v_2$, and $v_3$ satisfy $v^2=|v_1|^2+|v_2|^2+|v_3|^2=(174~\mathrm{GeV})^2$. After the mixing effects are absorbed into the effective Yukawa coupling matrices and normalized by the total electroweak VEV, the mass matrices of the charged leptons and quarks become
\begin{equation}
\begin{aligned}
    M_e=& \left(y^{d}_{\bar{5}}-3y^{d}_{\bar{45}}\right)^T v,\\
    M_d =&\left(y^{d}_{\bar{5}}+y^{d}_{\bar{45}}\right)v,\\
    M_u = & \; y_{5^\prime}^{u}v.
\end{aligned}
\end{equation}

For the neutrino sector, $NNf_1(Y)$ and $\bar{F}NH_5f_2(Y)$ correspond to the Majorana and Dirac neutrino interactions, respectively. The explicit Lagrangian reads
\begin{equation}
    -\mathcal{L}_{\nu}=a_1(NN)_{\mathbf{1}}Y^{(0)}_{\mathbf{1}}(\tau)+ a_2(NN)_{\mathbf{2}} Y^{(0)}_{\mathbf{2}}(\tau)+g_1(\bar{F}N)_{\mathbf{2}}
    Y^{(-4)}_{\mathbf{2}}(\tau)H_5 + g_2(\bar{F}N)_{\mathbf{3}} Y^{(-4)}_{\mathbf{3}}(\tau)H_5,
\end{equation}
where the first two terms determine the Majorana mass matrix $M_N$ of the right-handed neutrinos, while the latter two generate the neutrino Dirac mass matrix $M_D$ after electroweak symmetry breaking. With these conventions and after normalizing by the total electroweak VEV $v$, the Dirac and Majorana mass matrices of the neutrinos read
\begin{equation}
M_D =
    \begin{bmatrix}
        -2g_1Y^{(-4)}_{\mathbf{2},2}&g_2Y^{(-4)}_{\mathbf{3},2}&-g_2Y^{(-4)}_{\mathbf{3},3}\\
        g_2Y^{(-4)}_{\mathbf{3},2}&\sqrt{3}g_1Y^{(-4)}_{\mathbf{2},1}+g_2Y^{(-4)}_{\mathbf{3},1}&Y^{(-4)}_{\mathbf{2},2}\\
        -g_2Y^{(-4)}_{\mathbf{3},3}&g_1Y^{(-4)}_{\mathbf{2},2}&\sqrt{3}g_1Y^{(-4)}_{\mathbf{2},1}-g_2Y^{(-4)}_{\mathbf{3},1}
    \end{bmatrix}v,
\end{equation}
and
\begin{equation}
M_N =
    \begin{bmatrix}
        a_1Y^{(0)}_{\mathbf{1}}+2a_2Y^{(0)}_{\mathbf{2},1}&0&0\\
        0&\sqrt{3}a_2Y^{(0)}_{\mathbf{2},2}&a_1Y^{(0)}_{\mathbf{1}}-a_2Y^{(0)}_{\mathbf{2},1}\\
        0&a_1Y^{(0)}_{\mathbf{1}}-a_2Y^{(0)}_{\mathbf{2},1}&\sqrt{3}a_2Y^{(0)}_{\mathbf{2},2}
\end{bmatrix}.
\label{eq:mn}
\end{equation}
In the type-I seesaw mechanism, the light neutrino mass matrix is $M_\nu=-M_D M_N^{-1}M_D^T$.

In this model, field redefinitions allow $\alpha_1$, $\alpha_2$, $\alpha_3$, $\alpha'_1$, $\beta_3$, $a_1$, and $g_1$ to be taken as real, whereas the remaining parameters $\alpha'_2$, $\alpha'_3$, $\beta_1$, $\beta_2$, $a_2$, and $g_2$ are generically complex. For the numerical analysis, we introduce the following reparameterization to isolate the overall mass scales and reduce the number of independent dimensionless quantities:
\begin{align*}
&\tilde{\alpha}_1=\frac{\alpha_1}{\alpha_3}, \quad \tilde{\alpha}_2=\frac{\alpha_2}{\alpha_3}, \quad \tilde{\alpha}'_1=\frac{\alpha'_1}{\alpha_3}, \quad \tilde{\alpha}'_2=\frac{\alpha'_2}{\alpha_3}, \quad \tilde{\alpha}'_3=\frac{\alpha'_3}{\alpha_3}, \\
&\tilde{\beta}_1=\frac{\beta_1}{\beta_3}, \quad \tilde{\beta}_2=\frac{\beta_2}{\beta_3}, \quad \tilde{g}_2=\frac{g_2}{g_1}, \quad \tilde{a}_2=\frac{a_2}{a_1}.
\end{align*}
Among these, $\tilde{\alpha}_1$, $\tilde{\alpha}_2$, and $\tilde{\alpha}'_1$ are real, whereas the others are complex. With this parametrization, the down-type quark and charged-lepton mass matrices are $M_d=\hat{\alpha}_3 m_d$ and $M_e=\hat{\alpha}_3 m_e$ with $\hat{\alpha}_3=\alpha_3 v$, while the up-type quark mass matrix is $M_u=\hat{\beta}_3 m_u$ with $\hat{\beta}_3=\beta_3 v$. For the neutrino sector, $M_\nu=r m_\nu$, where the overall scale factor is $r=(g_1 v)^2/a_1$. These dimensionless parameters, together with the complex modulus $\tau$, therefore form the complete set of free parameters used in the model fitting.

\section{Fermion masses and mixing}
\label{sec:fit}
In this section, we perform a numerical and phenomenological analysis of the model to assess its viability in reproducing the observed fermion masses and mixing and to predict experimental parameters that have not yet been precisely measured. We take the parameters defined in Sec.~\ref{sec:model}, $\tilde{\alpha}_1,\dots,\tilde{\beta}_2,\tilde{g}_2,\tilde{a}_2$ and $\tau$, as inputs and scan the parameter space to identify the best-fit point. For all parameters except the modulus $\tau$, we scan their moduli in the range $(10^{-4},10^{4})$, and the phases of the complex parameters are varied in $(0,2\pi)$. These ranges are sufficiently broad to encompass all relevant cases while avoiding fine-tuning. The VEV $\langle\tau\rangle=\mathrm{Re}\,\tau+i\,\mathrm{Im}\,\tau$ determines the structure of the fermion mass matrices. We require $\tau$ to lie in its fundamental domain,
\begin{equation*}
    \left|\mathrm{Re}\,\tau\right|\le\frac12,\qquad \left|\tau\right|\ge 1. 
\end{equation*}

Building on this scan, we use the $\chi^2$ statistic to assess the fit quality of the model to the experimental data. For each scanned point, $\chi^2$ is computed, and the point that minimizes $\chi^2$ is identified as the best-fit point. The statistic is defined as
\begin{equation}
\chi^2=\sum_{i}\left(\frac{p_i-q_i}{\sigma_i}\right)^2,
\end{equation}
where $p_i$ is the model prediction for a given observable, $q_i$ is the corresponding experimental central value, and $\sigma_i$ is the associated $1\sigma$ uncertainty. A smaller $\chi^2$ indicates better agreement between the model and the data. The experimental constraints are summarized in Table~\ref{tab:input_data}. The parameter scan and the minimization of $\chi^2$ are performed with the \texttt{FlavorPy} package~\cite{FlavorPy}, a computational tool for flavor-model analysis.
\begin{table}[htbp]
\centering
\renewcommand{\arraystretch}{1.3}
\begin{tabular}{|c|c||@{\hspace{0.6cm}}c|c|}
\hline
\makecell{Parameters \\ (Charged Fermions)} & $\mu_i \pm 1\sigma$ & \makecell{Parameters \\ (Neutrinos)} & $\mu_i \pm 1\sigma$ \\
\hline
$m_e/m_\mu$ & $0.004737 \pm 0.000007$ & $\Delta m_{21}^2/10^{-5}\,\mathrm{eV}^2$ & $7.537_{-0.10}^{+0.094}$ \\
$m_\mu/m_\tau$ & $0.058823 \pm 0.000083$ & $\Delta m_{31}^2/10^{-3}\,\mathrm{eV}^2 (\text{NO})$ & $+2.511_{-0.020}^{+0.021}$ \\
$m_u/m_c$ & $0.001857 \pm 0.000629$ & $\Delta m_{32}^2/10^{-3}\,\mathrm{eV}^2 (\text{IO})$ & $-2.483_{-0.020}^{+0.020}$ \\
$m_c/m_t$ & $0.003194 \pm 0.000098$ & $\sin^2\theta_{12}^l$ & $0.3088_{-0.0066}^{+0.0067}$ \\
$m_d/m_s$ & $0.052827 \pm 0.005812$ & $\sin^2\theta_{23}^l (\text{NO})$ & $0.470_{-0.014}^{+0.017}$ \\
$m_s/m_b$ & $0.021710 \pm 0.001163$ & $\sin^2\theta_{23}^l (\text{IO})$ & $0.550_{-0.016}^{+0.013}$ \\
$\theta_{12}^q$ & $0.2274 \pm 0.00052$ & $\sin^2\theta_{13}^l (\text{NO})$ & $0.02248_{-0.00059}^{+0.00055}$ \\
$\theta_{13}^q$ & $0.0042 \pm 0.00017$ & $\sin^2\theta_{13}^l (\text{IO})$ & $0.02262_{-0.00056}^{+0.00057}$ \\
$\theta_{23}^q$ & $0.0485 \pm 0.00070$ & & \\
$\delta^q/^\circ$ & $69.16 \pm 3.09$ & & \\
\hline
\end{tabular}
\caption{Values of the charged-fermion mass ratios, quark mixing parameters, and neutrino oscillation observables used in the numerical fit. The charged-fermion parameters are evaluated at the GUT scale ($\mu=2\times10^{16}$ GeV) using the RGE running of Ref.~\cite{Babu:2016bmy}. The neutrino oscillation parameters are taken from the latest global fit (NuFit 6.1~\cite{Esteban:2024eli}).}
\label{tab:input_data}
\end{table}

We now present the results of our numerical fit. We scan the parameter space for the NO and IO cases separately and obtain the corresponding best-fit points, which are listed in Table~\ref{tab:bestfit_compare}. For the NO case, $\chi^2_{\min}=0.9190$. The leptonic sector, which includes the precisely measured neutrino oscillation parameters, is fitted very well, while the quark sector yields $\chi^2=0.9150$ and accounts for almost the entire total $\chi^2$. The largest contribution to $\chi^2$ comes from $m_u/m_c$, although it still lies within the $1\sigma$ experimental range.

The best-fit point predicts a sum of neutrino masses $\sum_i m_i = 0.06094\ \mathrm{eV}$. This value lies below the Planck upper bound $\sum m_i < 0.12\ \mathrm{eV}$ (95\% C.L.)~\cite{Vagnozzi:2017ovm,Planck:2018vyg} and also below the tightest bound from DESI BAO combined with CMB data in the $\Lambda$CDM framework, $\sum m_i < 0.072\ \mathrm{eV}$ (95\% C.L.)~\cite{DESI:2024mwx}. We note that this tightest bound is model-dependent. In extended cosmologies, such as the $w_0w_a$CDM model, the limit relaxes to approximately $0.177\ \mathrm{eV}$~\cite{Chebat:2025kes}, well above the model prediction. The best-fit point yields an effective electron neutrino mass $m_\beta = 0.009193\ \mathrm{eV}$ and an effective Majorana neutrino mass $m_{\beta\beta} = 0.002418\ \mathrm{eV}$. The former is well below the KATRIN upper limit $m_\beta < 0.45\ \mathrm{eV}$ (90\% C.L.)~\cite{Katrin:2024tvg}, while the latter is below the KamLAND-Zen range $m_{\beta\beta} < 0.028\text{–}0.122\ \mathrm{eV}$~\cite{KamLAND-Zen:2024eml}. In addition, the best-fit point predicts a Dirac CP phase $\delta_{\rm CP} = 1.723\pi$ and two Majorana phases $\eta_1 = 1.6743\pi$ and $\eta_2 = 1.8307\pi$. These Majorana phases currently have no direct experimental constraints and can only be probed indirectly, e.g.\ via neutrinoless double-beta decay ($0\nu\beta\beta$). These predictions therefore serve as targets for future experiments.
\begin{table}[!t]
\centering
\small
\renewcommand{\arraystretch}{1.12}
\setlength{\tabcolsep}{6pt}
\begin{tabular}{|c|c|c|c|}
\hline
\multicolumn{2}{|c|}{\textbf{Mass ordering}} & \textbf{NO} & \textbf{IO}\\
\hline
\multirow{12}{*}{\shortstack{Model \\ parameters}} 
& $\tilde{a}_2$ & $18.761\,e^{0.4784\pi i}$ & $4.8784\,e^{0.3626\pi i}$\\
& $\tilde{g}_2$ & $1.195\,e^{1.9237\pi i}$ & $1.8045\,e^{0.9095\pi i}$\\
& $\tilde{\alpha}'_2$ & $148.34\,e^{1.0541\pi i}$ & $3.8427\,e^{0.9983\pi i}$\\
& $\tilde{\alpha}'_3$ & $3419.9\,e^{1.7330\pi i}$ & $91.626\,e^{1.5294\pi i}$\\
& $\tilde{\beta}_1$ & $6.4052\times10^{-4}\,e^{1.8923\pi i}$ & $6.8092\times10^{-4}\,e^{0.5064\pi i}$\\
& $\tilde{\beta}_2$ & $0.11284\,e^{1.0769\pi i}$ & $0.14205\,e^{0.8053\pi i}$\\
\cline{2-4}
& $\tilde{\alpha}_{1}$ & $68.277$ & $1.386$\\
& $\tilde{\alpha}_{2}$ & $86.134$ & $5.9336$\\
& $\tilde{\alpha}'_1$ & $5.5956$ & $1.063$\\
& $r$ & $0.24604$ eV & $0.23391$ eV\\
& $\mathrm{Re}\, \tau $ & $0.33784$ & $-0.26190$\\
& $\mathrm{Im}\, \tau $ & $0.99399$ & $0.98757$\\
\hline
\multirow{24}{*}{\shortstack{Predicted \\ quantities}} 
& $m_e/m_\mu$ & $4.7370\times10^{-3}$ & $4.7370\times10^{-3}$\\
& $m_\mu/m_\tau$ & $5.8820\times10^{-2}$ & $5.8820\times10^{-2}$\\
& $\sin^2\theta_{12}$ & $0.3087$ & $0.3088$\\
& $\sin^2\theta_{13}$ & $0.02248$ & $0.02262$\\
& $\sin^2\theta_{23}$ & $0.4709$ & $0.5497$\\
& $\delta_{CP}/\pi$ & $1.7233$ & $0.7703$\\
& $m_1$ & $1.9054\times10^{-3}$ eV & $4.9116\times10^{-2}$ eV\\
& $m_2$ & $8.8887\times10^{-3}$ eV & $4.9878\times10^{-2}$ eV\\
& $m_3$ & $5.0143\times10^{-2}$ eV & $2.1908\times10^{-3}$ eV\\
& $\sum_i m_i$ & $0.06094$ eV & $0.10118$ eV\\
& $\Delta m^2_{21}$ & $7.5379\times10^{-5}$ eV$^2$ & $7.5371\times10^{-5}$ eV$^2$\\
& $\Delta m^2_{31}$ & $2.5107\times10^{-3}$ eV$^2$ & $-2.4830\times10^{-3}$ eV$^2$\\
& $m_\beta$ & $9.1933\times10^{-3}$ eV & $4.9825\times10^{-2}$ eV\\
& $m_{\beta\beta}$ & $2.4179\times10^{-3}$ eV & $2.4644\times10^{-2}$ eV\\
& $\eta_1/\pi$ & $1.6743$ & $0.4652$\\
& $\eta_2/\pi$ & $1.8307$ & $1.0867$\\
\cline{2-4}
& $m_u/m_c$ & $2.2258\times10^{-3}$ & $2.3537\times10^{-3}$\\
& $m_c/m_t$ & $3.1522\times10^{-3}$ & $3.1626\times10^{-3}$\\
& $m_d/m_s$ & $5.3498\times10^{-2}$ & $5.1920\times10^{-2}$\\
& $m_s/m_b$ & $2.2054\times10^{-2}$ & $2.1513\times10^{-2}$\\
& $\theta^{q}_{12}$ & $0.22740$ & $0.22739$\\
& $\theta^{q}_{13}$ & $4.1859\times10^{-3}$ & $4.2254\times10^{-3}$\\
& $\theta^{q}_{23}$ & $4.8771\times10^{-2}$ & $4.8722\times10^{-2}$\\
& $\delta^q$ & $70.281^\circ$ & $70.648^\circ$\\
\hline
\multirow{3}{*}{\shortstack{Goodness of fit}} 
& $\chi^2_{\rm lepton}$ & $4.03\times10^{-3}$ & $4.84\times10^{-4}$\\
& $\chi^2_{\rm quark}$ & $0.9150$ & $1.1344$\\
& $\chi^2_{\rm total}$ & $0.9190$ & $1.1349$\\
\hline
\end{tabular}
\caption{Input parameters and predicted observables at the best-fit points of the modular $S'_4$ model for normal (NO) and inverted (IO) neutrino mass orderings.}
\label{tab:bestfit_compare}
\end{table}

For the IO case, we obtain $\chi^2_{\min}=1.1349$. The leptonic sector is again fitted very well, while the quark sector yields $\chi^2=1.1344$ and again dominates the total $\chi^2$. The best-fit point predicts a sum of neutrino masses $\sum m_i = 0.10118\ \mathrm{eV}$, which lies below the Planck cosmological bound $\sum m_i < 0.12\ \mathrm{eV}$~\cite{Vagnozzi:2017ovm,Planck:2018vyg}. Since the neutrino mass spectrum is more degenerate in the IO case, $m_\beta$ and $m_{\beta\beta}$ are larger than in the NO case. The best-fit point gives $m_\beta = 0.049825\ \mathrm{eV}$ and $m_{\beta\beta} = 0.024644\ \mathrm{eV}$, still below the KATRIN~\cite{Katrin:2024tvg} and KamLAND-Zen~\cite{KamLAND-Zen:2024eml} limits. The predicted phases are $\delta_{\rm CP} = 0.7703\pi$, $\eta_1 = 0.4652\pi$, and $\eta_2 = 1.0867\pi$.

Comparing the two cases, we find $\Delta\chi^2=\chi^2_{\min}(\mathrm{NO})-\chi^2_{\min}(\mathrm{IO})=-0.2159$, indicating that the model fits the NO case slightly better. In both cases, the total $\chi^2$ is dominated by the quark sector, while the leptonic sector is fitted very well. To further assess the model against the experimental data, we perform a Markov Chain Monte Carlo (MCMC) scan around each best-fit point with the \texttt{FlavorPy} package, in order to explore the parameter uncertainties.

\begin{figure}[htbp]
    \centering
    \includegraphics[width=\textwidth]{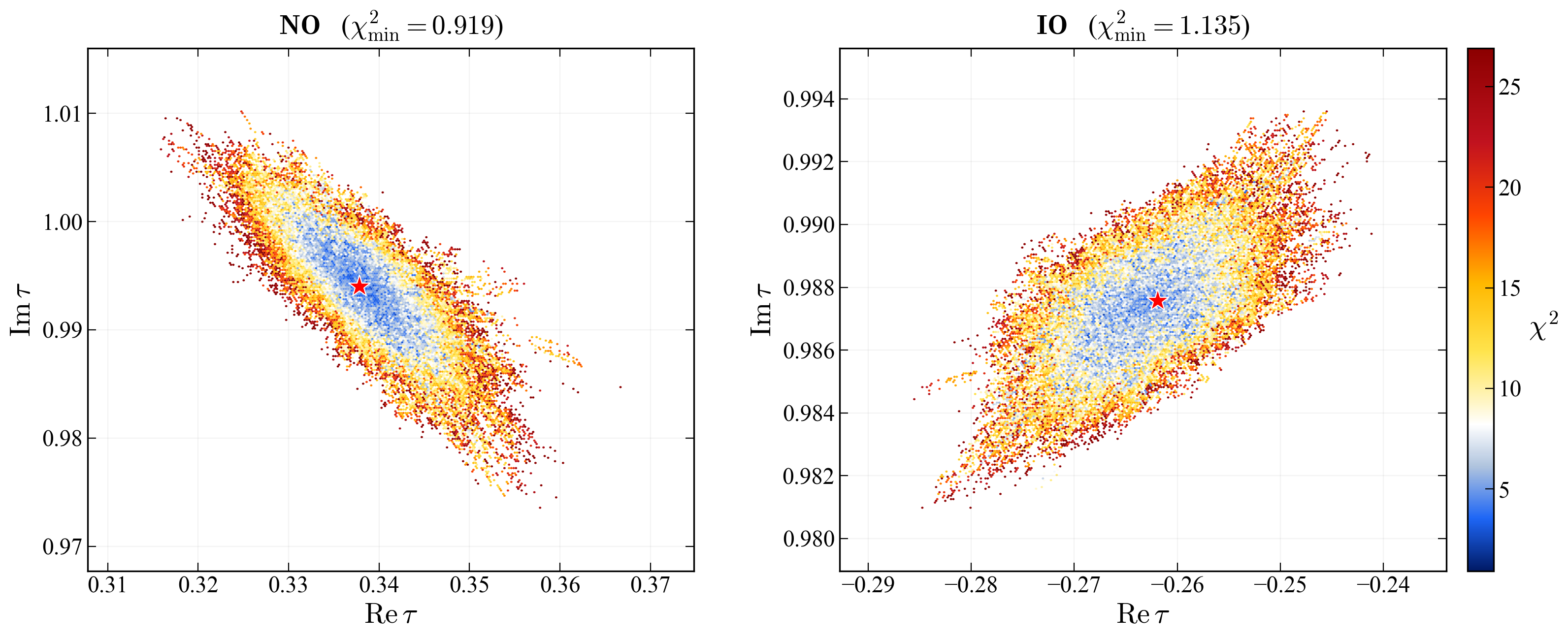}
    \caption{The distribution of the modulus $\tau$ in the NO (left) and IO (right) cases. The red star marks the best-fit point, and the color indicates the $\chi^2$ value, with darker red corresponding to larger $\chi^2$.}
    \label{fig:tau}
\end{figure}

Figure~\ref{fig:tau} shows the distribution of the modulus $\tau$ in the NO (left panel) and IO (right panel) cases. The color scale indicates the $\chi^2$ value.  In the NO case, the best-fit point lies at $\mathrm{Re}\,\tau=0.3378$ and $\mathrm{Im}\,\tau\approx0.9940$, and the scan points mainly populate the ranges $\mathrm{Re}\,\tau\in[0.312,0.368]$ and $\mathrm{Im}\,\tau\in[0.972,1.01]$, with a negative correlation between $\mathrm{Re}\,\tau$ and $\mathrm{Im}\,\tau$. In the IO case, the best-fit point shifts to $\mathrm{Re}\,\tau\approx-0.2619$ and $\mathrm{Im}\,\tau\approx0.9876$, with the scan points distributed in $\mathrm{Re}\,\tau\in[-0.292,-0.240]$ and $\mathrm{Im}\,\tau\in[0.983,0.994]$, showing a positive correlation between $\mathrm{Re}\,\tau$ and $\mathrm{Im}\,\tau$.
\begin{figure}[ht]
    \centering
    \includegraphics[width=\textwidth]{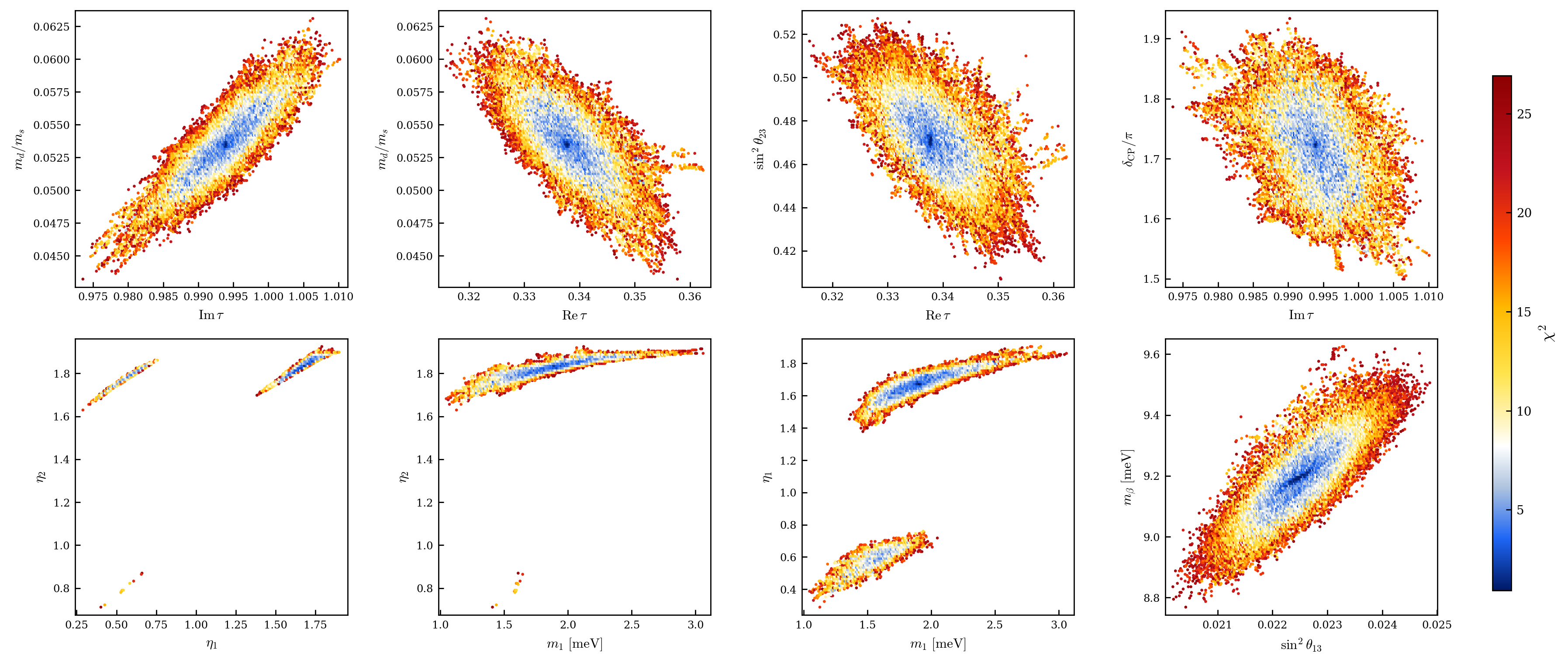}
    \caption{Two-dimensional posterior correlations between the modulus $\tau$ and the physical quantities $m_d/m_s$, $\sin^2\theta_{23}$, $\delta_{\mathrm{CP}}/\pi$, $\eta_1$, $\eta_2$, $m_1$, $m_\beta$, and $\sin^2\theta_{13}$ in the region $\mathrm{Re}\, \tau >0$, corresponding to the NO case.}
    \label{fig:NO_correlation_2d}
\end{figure}

Figure~\ref{fig:NO_correlation_2d} further shows the correlations between $\tau$ and the physical quantities in the NO case. The ratio $m_d/m_s$ is positively correlated with $\mathrm{Im}\,\tau$ and negatively correlated with $\mathrm{Re}\,\tau$. The Dirac CP phase $\delta_{\mathrm{CP}}$, in the sense of the overall shape, decreases as $\mathrm{Im}\,\tau$ increases. The Majorana phase $\eta_1$ and the lightest neutrino mass $m_1$ exhibit a pronounced piecewise positive correlation, while $m_3$ and $\sin^2\theta_{13}$ show a positive correlation.
\begin{figure}[htbp]
    \centering
    \includegraphics[width=\textwidth]{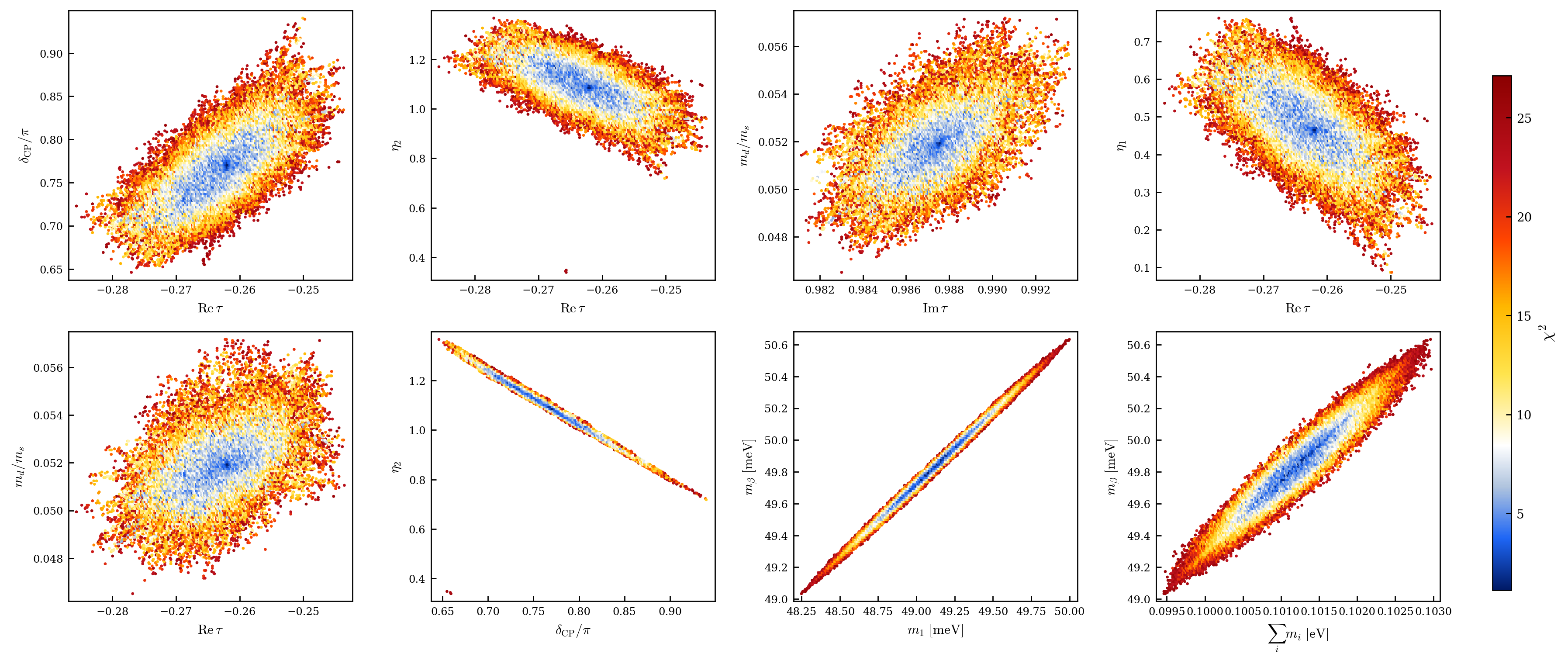}
    \caption{Two-dimensional posterior correlations in the IO case, shown in the region $\mathrm{Re}\,\tau<0$. The panels display the correlations of $\delta_{\mathrm{CP}}$, $\eta_2$ and $m_d/m_s$ with $\mathrm{Re}\,\tau$, the ratio $m_d/m_s$ with $\mathrm{Im}\,\tau$, as well as the strong correlations between $\eta_2$ and $\delta_{\rm CP}$, and between $m_{\beta}$ and $m_1$, and the mild positive correlation between $m_{\beta}$ and the sum of neutrino masses $\sum_i m_i$.}
    \label{fig:IO_correlation_2d}
\end{figure}

Figure~\ref{fig:IO_correlation_2d} shows the parameter correlations in the region $\mathrm{Re}\,\tau<0$, corresponding to the IO case. Here $\delta_{\mathrm{CP}}$ is positively correlated with $\mathrm{Re}\,\tau$, while $\eta_1$ and $\eta_2$ are negatively correlated with $\mathrm{Re}\,\tau$. In particular, $\eta_2$ and $\delta_{\rm CP}$ (approximately $0.65\pi\text{–}0.95\pi$), as well as $m_\beta$ and $m_1$ (approximately $48.2\text{–}50.0$ meV), exhibit strong correlations. This reflects the stringent constraint that the model imposes on the neutrino mass spectrum in the IO case.
\begin{figure}[htbp]                  
  \centering
    \includegraphics[width=\textwidth]{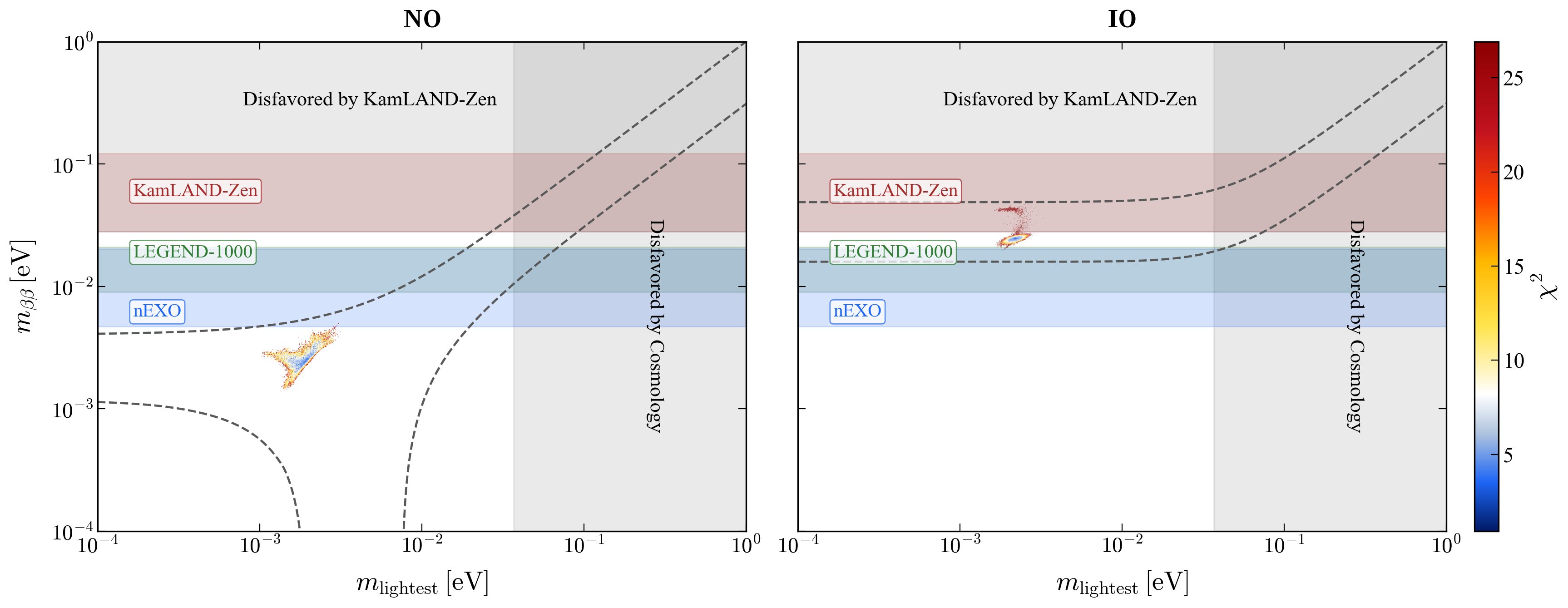} 
    \caption{Model predictions for the effective Majorana mass $m_{\beta\beta}$ versus the lightest neutrino mass $m_{\mathrm{lightest}}$ in the NO (left) and IO (right) cases.  The color of the scan points indicates the $\chi^2$ value. The horizontal bands represent experimental sensitivities: the KamLAND-Zen exclusion limit (brownish-red band, $m_{\beta\beta} < 0.028$ - $0.122$ eV~\cite{KamLAND-Zen:2024eml}), and the target sensitivities of next-generation $0\nu\beta\beta$ experiments, LEGEND-1000 (green band, $0.009 $ - $0.021$ eV~\cite{LEGEND:2021bnm}) and nEXO (blue band, $0.0047$-$0.0203$ eV~\cite{nEXO:2021ujk}). The gray region on the right corresponds to the cosmological bound~\cite{GAMBITCosmologyWorkgroup:2020rmf}.}
  \label{fig:mbb}
\end{figure}

Figure~\ref{fig:mbb} displays the model predictions for the effective Majorana neutrino mass $m_{\beta\beta}$ as a function of the lightest neutrino mass $m_{\mathrm{lightest}}$. In the NO case (left panel), the scan points are concentrated in the range $m_{\mathrm{lightest}}\in[1.03\times10^{-3},\,3.14\times10^{-3}]$ eV, with corresponding $m_{\beta\beta}$ predictions of approximately $1.40\times10^{-3}$–$4.99\times10^{-3}$ eV. This region lies well below the current KamLAND-Zen exclusion limit and entirely within the cosmological bound. The predicted range partially overlaps with the target sensitivity windows of nEXO and LEGEND-1000, while the remainder lies below their reach. In the IO case (right panel), the lower bound on the lightest neutrino mass is raised, so the predicted $m_{\beta\beta}$ shifts upward to approximately $0.02$ eV. The scan points then lie very close to, and partially touch, the lower end of the KamLAND-Zen exclusion range ($0.028$ eV). This indicates that, in the IO case, the predicted signal lies at the edge of current experimental sensitivity.
\begin{figure}[ht]                  
  \centering
    \includegraphics[width=\textwidth]{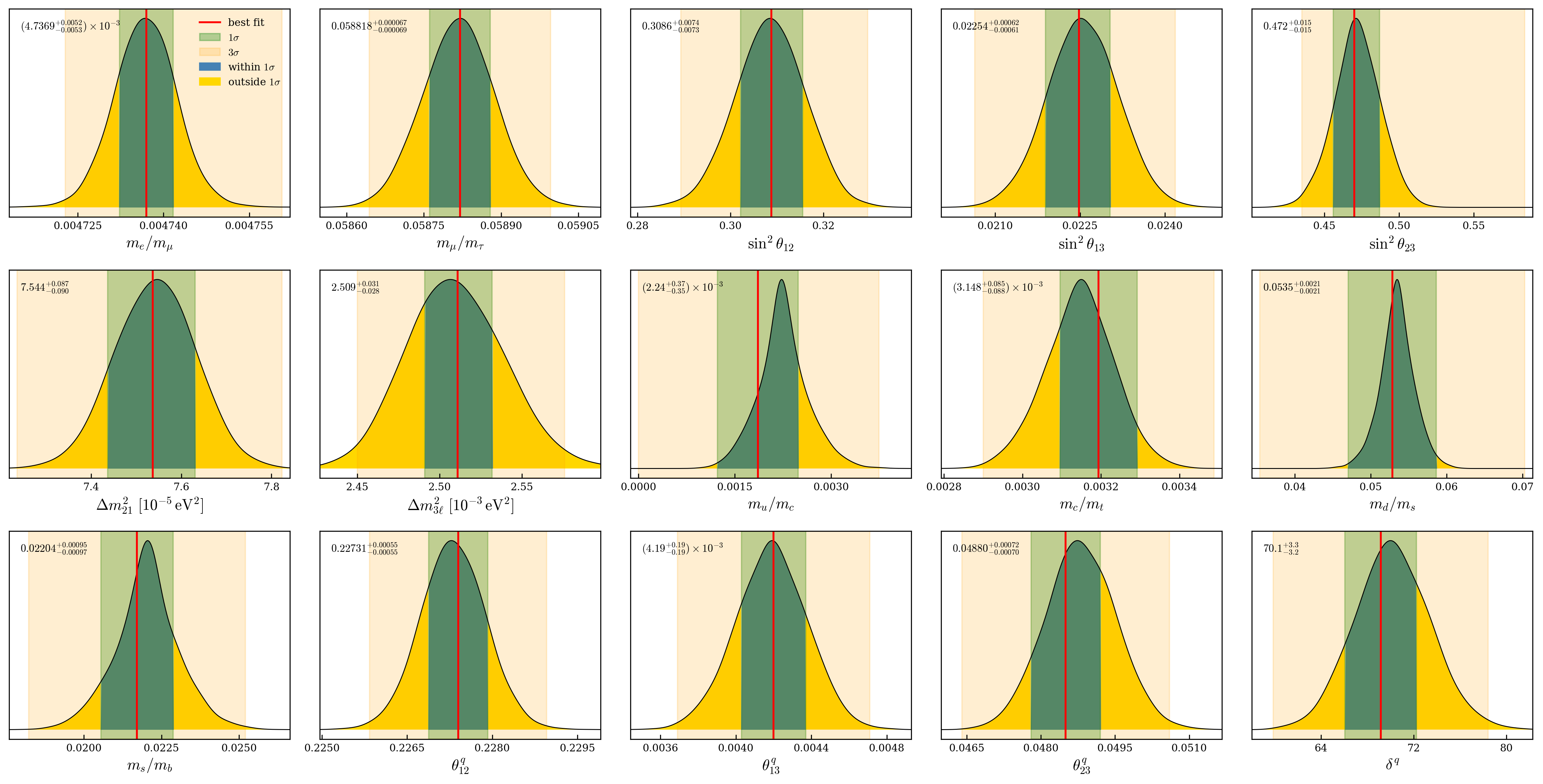} 
    \caption{One-dimensional marginalized posterior distributions of the fifteen fitted observables in the NO case, obtained from the MCMC chains. In each panel, the red vertical line denotes the experimental central value, and the green and orange shaded bands indicate the $1\sigma$ and $3\sigma$ allowed ranges, respectively. The area under each posterior is colored blue inside the $1\sigma$ interval and yellow outside. Each panel is annotated with the posterior median and the asymmetric uncertainties given by the $68\%$ credible interval. }
  \label{fig:NO_obs_kde}
\end{figure}

\begin{figure}[htbp]                  
  \centering
  \includegraphics[width=\textwidth]{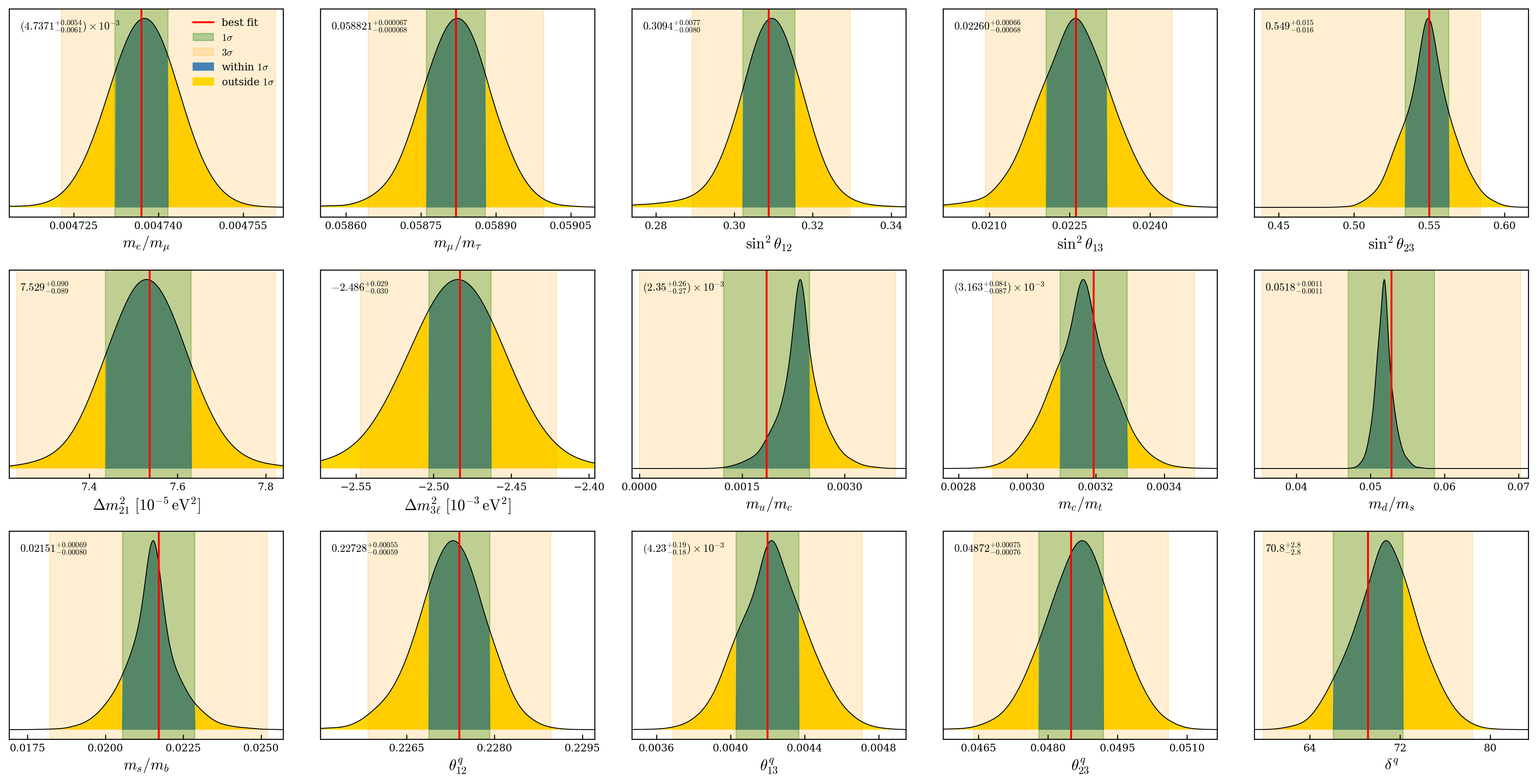} 
    \caption{One-dimensional marginalized posterior distributions of the fifteen fitted observables in the IO case.  The panel conventions are the same as in Fig.~\ref{fig:NO_obs_kde}.}
  \label{fig:IO_obs_kde}
\end{figure}

Figures~\ref{fig:NO_obs_kde} and~\ref{fig:IO_obs_kde} show the one-dimensional marginalized posterior distributions of the fifteen fitted observables in the NO and IO cases, respectively. The posteriors are obtained from the MCMC chains. In each panel, the red vertical line denotes the experimental central value, and the green and orange shaded bands indicate the $1\sigma$ and $3\sigma$ experimental allowed ranges, respectively. The part of the posterior inside the $1\sigma$ interval is colored blue, and the part outside is colored yellow. All fifteen posteriors are sharply concentrated within the experimental ranges. The yellow regions reflect only the finite width of the posteriors and do not indicate any systematic deviation. The values quoted in each panel are the posterior median and the $68\%$ credible interval given by the $16\%$ and $84\%$ quantiles. In both cases, the posterior median of $m_u/m_c$ deviates slightly from the experimental central value but remains within the $1\sigma$ interval.

\begin{figure}[ht]
  \centering
  \includegraphics[width=\linewidth]{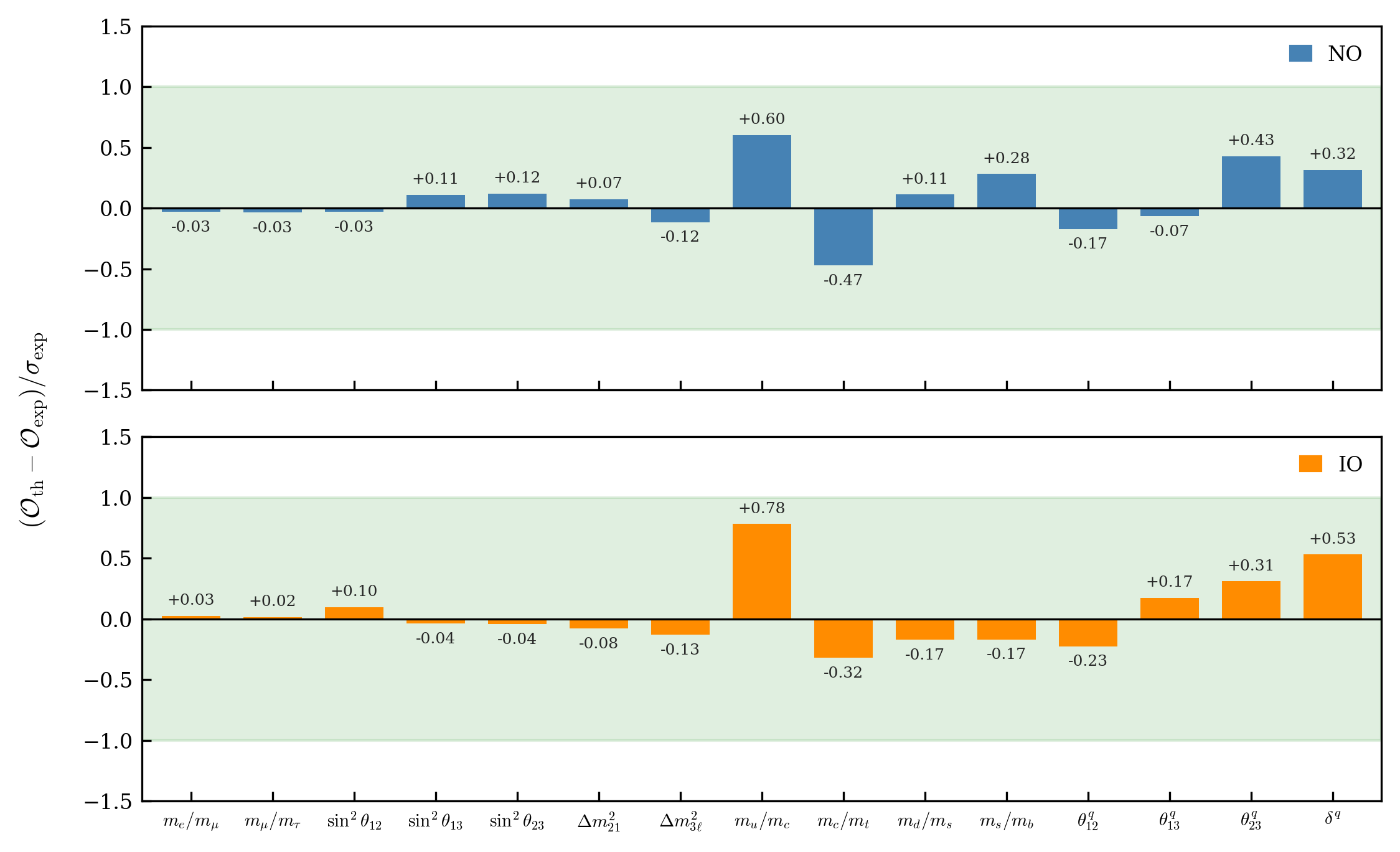}
    \caption{Model-predicted Pull values $\mathcal{P}=(\mathcal{O}_{\mathrm{th}}-\mathcal{O}_{\mathrm{exp}})/\sigma_{\mathrm{exp}}$ for the fifteen fitted observables, where $\sigma_{\mathrm{exp}}$ is the experimental uncertainty. The vertical error bars represent the predicted posterior uncertainties in units of $\sigma_{\mathrm{exp}}$. Blue and orange bars correspond to the NO and IO cases, respectively. The green horizontal bands indicate the $\pm1\sigma$ regions. All fifteen Pull values satisfy $|\mathcal{P}|<0.8$.}
  \label{fig:pull_NO_IO}
\end{figure}

Figure~\ref{fig:pull_NO_IO} quantifies the agreement between the model and the data through the Pull values $\mathcal{P}=(\mathcal{O}_{\mathrm{th}}-\mathcal{O}_{\mathrm{exp}})/\sigma_{\mathrm{exp}}$ for the fifteen fitted observables, where $\sigma_{\mathrm{exp}}$ is the experimental uncertainty. The vertical error bars represent the predicted posterior uncertainties in units of $\sigma_{\mathrm{exp}}$. The blue and orange bars correspond to the NO and IO cases, respectively. All fifteen absolute Pull values satisfy $|\mathcal{P}|<0.8$. The largest deviations appear in the quark sector, most notably for $m_u/m_c$ ($+0.60$ for NO and $+0.78$ for IO), followed by $m_c/m_t$ ($-0.47$) in the NO case and $\delta^q$ ($+0.53$) in the IO case. In contrast, the leptonic observables agree very well with the experimental central values, with deviations below $0.15\sigma$, and the remaining quark parameters also deviate by at most about $0.3\sigma$. We therefore conclude that the model reproduces all fitted observables well at the $1\sigma$ level, with no statistically significant tension in either the leptonic or the quark sector.

\begin{figure}[ht]
  \centering
  \includegraphics[width=\textwidth]{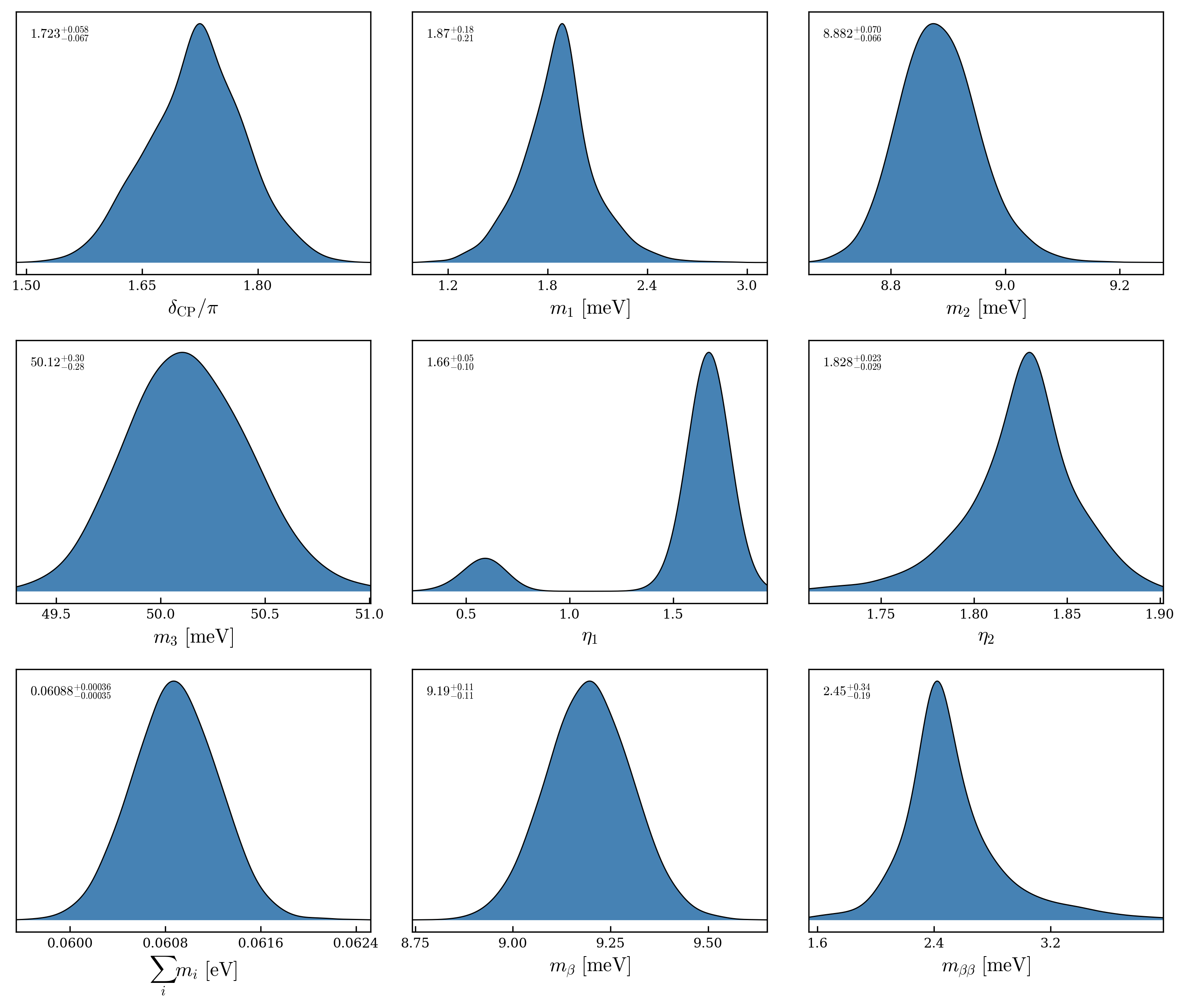} 
    \caption{One-dimensional posterior distributions of selected neutrino parameters in the NO case, including the CP phase $\delta_{\mathrm{CP}}$, the neutrino masses $m_1$, $m_2$, and $m_3$, the Majorana phases $\eta_1$ and $\eta_2$, the sum of neutrino masses $\sum_i m_i$, and the effective masses $m_\beta$ and $m_{\beta\beta}$.}
  \label{fig:NO_pre_kde}
\end{figure}

Figure~\ref{fig:NO_pre_kde} illustrates the one-dimensional posterior distributions of selected neutrino parameters in the NO case: the neutrino masses $m_i$, the Majorana phases $\eta_1$ and $\eta_2$, the effective masses $m_\beta$ and $m_{\beta\beta}$, and the sum of neutrino masses $\sum_i m_i$. The predicted mass spectrum exhibits the typical NO pattern, with medians $m_1=1.87$ meV, $m_2=8.882$ meV, and $m_3=50.12$ meV. The CP phase is concentrated around $\delta_{\mathrm{CP}}\simeq1.723$. The medians of the sum of neutrino masses, the effective electron neutrino mass, and the effective Majorana mass are $\sum_i m_i=0.06088$ eV, $m_\beta=9.19$ meV, and $m_{\beta\beta}=2.45$ meV, respectively. Most posteriors are single-peaked and approximately Gaussian. The Majorana phases, however, exhibit distinct structures: $\eta_1$ displays a bimodal distribution, with the main peak at $1.66\pi$ and a secondary peak near $0.5\pi$, whereas $\eta_2$ is predominantly single-peaked, centered at $1.828\pi$ with a long tail~\footnote{Although the MCMC samples contain a very weak probability distribution for $\eta_2$ around $0.8\pi$ (visible in Fig.~\ref{fig:NO_correlation_2d}), it does not appear in this one-dimensional plot because of its extremely small amplitude.}.
\begin{figure}[ht]
  \centering
  \includegraphics[width=\textwidth]{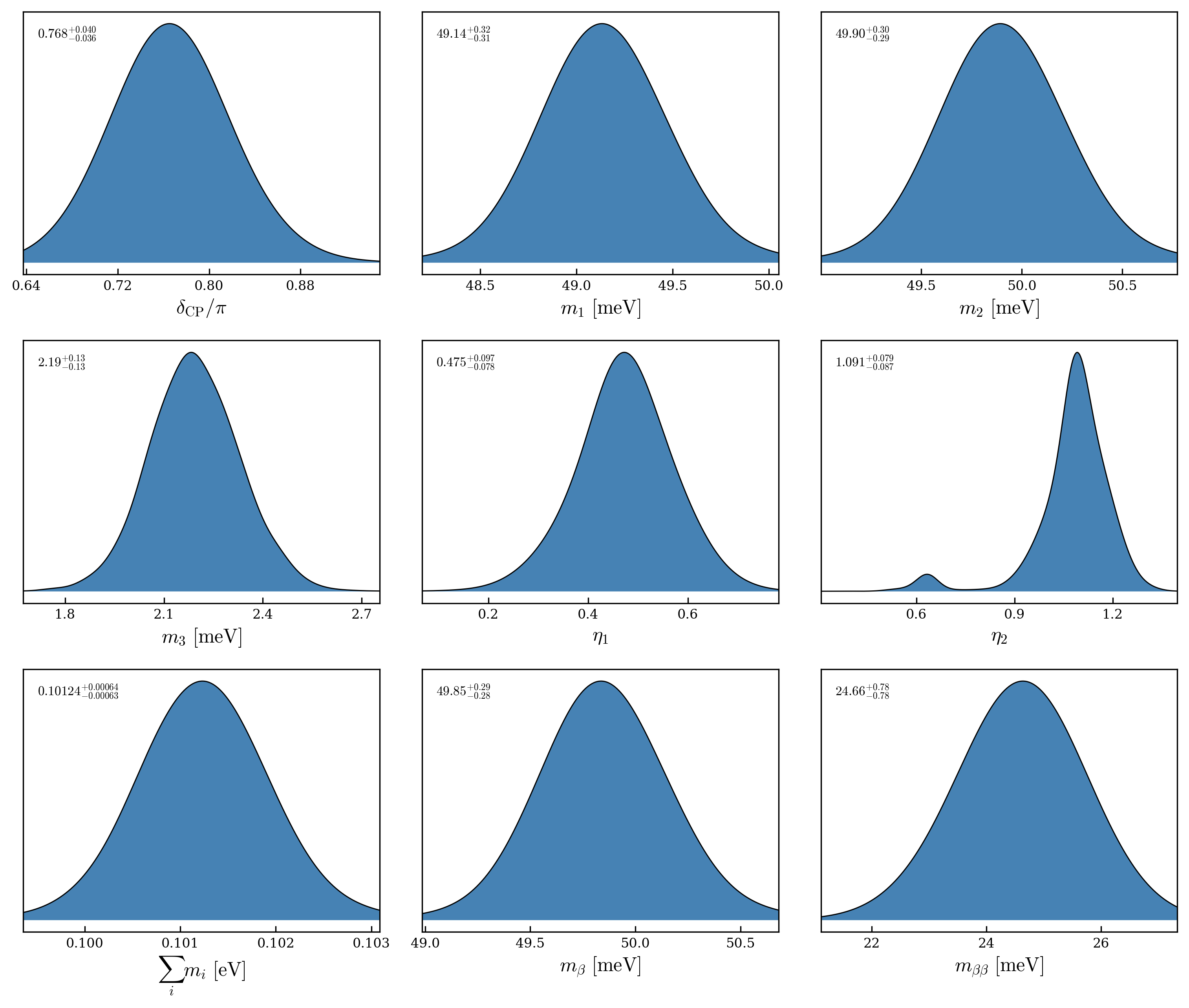} 
  \caption{One-dimensional posterior distributions of selected neutrino parameters in the IO case. The physical meaning of each parameter is the same as in Fig.~\ref{fig:NO_pre_kde}.}
  \label{fig:IO_pre_kde}
\end{figure}

Figure~\ref{fig:IO_pre_kde} shows the corresponding posterior distributions in the IO case. Here the mass spectrum exhibits the typical IO pattern, where two masses $m_1$ and $m_2$ are highly degenerate, with medians $49.14$ meV and $49.90$ meV, while $m_3$ is lighter at $2.19$ meV. The CP phase is concentrated around $\delta_{\mathrm{CP}}\simeq 0.768\pi$, and the medians of $\sum_i m_i$, $m_\beta$, and $m_{\beta\beta}$ are $0.10124$ eV, $49.85$ meV, and $24.66$ meV, respectively. Unlike the NO case, it is $\eta_2$ that is bimodal in the IO case, with the main peak at $1.091\pi$ and a secondary peak near $0.6\pi$, whereas $\eta_1$ is single-peaked. The model therefore provides distinct theoretical predictions for these yet-unmeasured parameters. These differences not only help test the model in future experiments but also offer a potential handle for determining the neutrino mass ordering.

\section{Thermal leptogenesis}
\label{sec:tlep}

In the Sec.~\ref{sec:fit}, we presented a systematic phenomenological analysis of the model and demonstrated its ability to reproduce the observed fermion masses and mixing. Since neutrino masses are generated through the type-I seesaw mechanism, the model also provides a natural framework for studying thermal leptogenesis. In this section, we investigate whether the model can account for the observed baryon asymmetry of the Universe.

We first redefine the neutrino mass matrices so that the physical mass matrices take the form
\begin{equation}
  M_N^{\rm phys}=b\,M_N,\qquad M_D^{\rm phys}=a\,M_D,\qquad
  a=\sqrt{b\,r},
  \label{eq:scales}
\end{equation}
and the light-neutrino mass matrix becomes
\begin{equation}
  m_\nu=-(M_D^{\rm phys})(M_N^{\rm phys})^{-1}(M_D^{\rm phys})^{T}
      =-r\,M_D\,M_N^{-1}M_D^{T}.
  \label{eq:seesaw}
\end{equation}
Here $a$ and $b$ are scale factors, and $r\equiv a^2/b$ is fixed by the fit described in the Sec.~\ref{sec:fit}. The scale factor $b$ sets the overall scale of the heavy neutrinos, while $r$ is determined by the fit. Consequently, for a given benchmark point, $b$ is the only free parameter. For the heavy-neutrino mass matrix in Eq.~\eqref{eq:mn}, the mass eigenvalues are
\begin{align}
    \lambda_1 &= b\left| 1 + 2\tilde{a}_2 Y_{\mathbf{2},1}^{(0)} \right|, \label{eq:mass1} \\
    \lambda_2 &= b\left| 1 - \tilde{a}_2 Y_{\mathbf{2},1}^{(0)} + \sqrt{3}\tilde{a}_2 Y_{\mathbf{2},2}^{(0)} \right|, \label{eq:mass2} \\
    \lambda_3 &= b\left| 1 - \tilde{a}_2 Y_{\mathbf{2},1}^{(0)} - \sqrt{3}\tilde{a}_2Y_{\mathbf{2},2}^{(0)} \right|. \label{eq:mass3}
\end{align}
The overall scale of the heavy-neutrino spectrum is therefore set entirely by $b$, while the ratios of adjacent masses, such as $\lambda_2/\lambda_1$ and $\lambda_3/\lambda_2$, depend only on the dimensionless parameter $\tilde{a}_2$ and the modular forms $Y$. Varying $b$ thus shifts the heavy-neutrino spectrum as a whole without changing its internal hierarchy. And it is the only parameter that governs the dynamics of thermal leptogenesis, including washout effects and decay rates.

To obtain the physical heavy-neutrino states, we work in the mass basis in which $M_N$ is diagonalized through the Takagi decomposition,
\begin{equation}
  U_N^{T}M_N U_N=\operatorname{diag}(M^U_1,M^U_2,M^U_3),\qquad M^U_1<M^U_2<M^U_3 .
  \label{eq:takagi}
\end{equation}
In this basis, the Yukawa couplings and the heavy masses are
\begin{equation}
  h=\frac{aM_D}{v} U_N,\qquad M_i=b\,M^U_i.
  \label{eq:hM}
\end{equation}
Here the Higgs VEV is $v=174\ \mathrm{GeV}$. In the early Universe, the out-of-equilibrium, CP-violating decays of these three heavy neutrinos violate lepton number and generate a $B-L$ asymmetry, which is subsequently converted into a baryon asymmetry by electroweak sphaleron processes. In the following, we analyze the evolution of the lepton asymmetry with the density matrix equations and numerically compute the final baryon asymmetry.

\subsection{Density matrix equations}
\label{subsec:densitymatrxeq}

In the framework of thermal leptogenesis, the production and decay of the three heavy neutrinos $N_i$ and the evolution of the lepton asymmetry are described by the Boltzmann equations. We use the density matrix formalism to track the asymmetries in flavor space and the coherence between different flavors. We define the dimensionless variables
\begin{equation}
    z_i\equiv\frac{M_i}{T}=\frac{M_i}{M_1}z,\qquad
    \widetilde m_i=\frac{(h^\dagger h)_{ii}v^2}{M_i},\qquad
    K_i=\frac{\widetilde m_i}{m_*}.
  \label{eq:zK}
\end{equation}
Here $T$ is the temperature of the cosmic plasma and $M_i$ is the mass of the heavy neutrino $N_i$. The variable $z=M_1/T$ characterizes the cosmic temperature, while $z_i=M_i/T$ is the same quantity scaled by the mass ratio $M_i/M_1$. In the definition of the effective neutrino mass $\widetilde m_i$, $h$ is the neutrino Yukawa coupling matrix and $v$ is the electroweak Higgs VEV. The effective neutrino mass $\widetilde m_i$ quantifies the coupling strength between the heavy neutrino $N_i$ and the lepton doublet and therefore the impact of its decay on the lepton asymmetry. The washout parameter $K_i$ measures the competition between the heavy-neutrino decays and the cosmic expansion, where $m_*\simeq 1.08\times10^{-3}$ eV is the characteristic neutrino mass scale, determined by the Hubble expansion rate and the heavy-neutrino decay rate during the radiation-dominated era. For $K_i\gg1$, decays and inverse decays are fast compared with the cosmic expansion and the system is in the strong washout regime. For $K_i\ll1$, on the other hand, these processes are too slow to maintain the heavy neutrinos in thermal equilibrium and the system is in the weak washout regime.

The abundance of the heavy neutrino $N_i$ is denoted by $N_{N_i}$, and its evolution equation is
\begin{equation}
    \frac{dN_{N_i}}{dz}=-D_i\left(N_{N_i}-N_{N_i}^{\rm eq}\right),
    \label{eq:NN_Boltzmann}
\end{equation}
where $N_{N_i}^{\rm eq}$ is the equilibrium abundance of the heavy neutrino and $D_i$ is a dimensionless function characterizing the decay and inverse decay processes. The explicit expressions for these quantities are~\cite{BuchmullerDiBariPlumacher2005}
\begin{align}
  N_{N_i}^{\rm eq}=& \frac38 z_i^2 K_2(z_i),\label{eq:Neq}\\  
  D_i= & K_i z\left(\frac{M_i}{M_1}\right)^2\frac{K_1(z_i)}{K_2(z_i)},
  \label{eq:Di}
\end{align}
where $K_n(z_i)$ is the modified Bessel function of the second kind. The lepton asymmetry is described by the density matrix $N_{\alpha\beta}=n_{\Delta_{\alpha\beta}}/n_\gamma$ in the flavor basis $(e,\mu,\tau)$, where $n_{\Delta_{\alpha\beta}}$ is the lepton asymmetry number density. The diagonal elements of $N_{\alpha\beta}$ describe the asymmetries of the individual lepton flavors, while the off-diagonal elements describe the quantum coherence between different flavors. 

The evolution equation of this density matrix is
\begin{equation}
  \frac{d N_{\alpha\beta}}{dz}=\sum_i\left[\varepsilon^{(i)}_{\alpha\beta}D_i\left(N_{N_i}-N_{N_i}^{\rm eq}\right)-\frac12 W_i\left \{P^{(i)},N\right \} _{\alpha\beta}\right] -\frac{\Gamma_\alpha+\Gamma_\beta}{Hz}\,N_{\alpha\beta}\,(1-\delta_{\alpha\beta}).
  \label{eq:dmb3}
\end{equation}
The right-hand side of Eq.~\eqref{eq:dmb3} contains three physical contributions. The first term is the CP-violating source of the lepton asymmetry, where $\varepsilon_{\alpha\beta}^{(i)}$ is the CP asymmetry matrix. The second term describes the washout of the generated asymmetry by inverse decays, where $W_i$ is the washout function and $P^{(i)}$ is the flavor projection matrix. This term acts on both the diagonal and off-diagonal elements of the density matrix through an anticommutator structure. The third term accounts for the decoherence induced by the charged-lepton Yukawa interactions, which acts only on the off-diagonal elements, where $\Gamma_\alpha$ is the Yukawa interaction rate of the charged lepton flavor $\alpha$ and $\delta_{\alpha\beta}$ is the Kronecker delta function. For numerical calculations, this interaction rate is approximated by $\Gamma_\alpha=C\,m_\alpha^2 T/v^2$ (with $m_\alpha$ the charged-lepton mass and the coefficient $C\sim 8\times10^{-3}$~\cite{GranelliEtAl2021,Ding2022}), while the Hubble expansion rate is $H=1.66\sqrt{g_*}T^2/M_{\rm Pl}$. This density matrix treatment allows us to track the coherence among different lepton flavors and its evolution with the expanding Universe~\cite{Stodolsky1987,Sigl1993,Blanchet2007,DeSimone2007,Beneke2010,Anisimov2011,BlanchetEtAl2013}.

To solve the evolution Eq.~\eqref{eq:dmb3}, we need to specify the microscopic expressions for each term on the right-hand side. For the first term, which determines the CP-violating source, the flavored CP asymmetry matrix is obtained at one-loop order as~\cite{CoviRouletVissani1996,Pilaftsis1997,Moffat2019}
\begin{align}
  \varepsilon^{(i)}_{\alpha\beta}=\frac{3}{32\pi(h^\dagger h)_{ii}}
  \sum_{j\neq i}\Bigg\{&
    \mathrm{i}\Big[(h^\dagger h)_{ji}h_{\alpha i}h_{\beta j}^\ast
      -(h^\dagger h)_{ij}h_{\alpha j}h_{\beta i}^\ast\Big]\,
      \frac{M_i}{M_1}\,f_1\!\left(x_{ji}\right)\nonumber\\[2pt]
   &+\mathrm{i}\,f_2\!\left(x_{ji}\right)
     \Big[h_{\alpha i}h_{\beta j}^\ast(h^\dagger h)_{ij}
       -h_{\beta i}^\ast h_{\alpha j}(h^\dagger h)_{ji}\Big]\Bigg\},
  \label{eq:eps}
\end{align}
where the first term corresponds to the vertex contribution, the second to the self-energy contribution, and $x_{ji}\equiv M_j^2/M_i^2$ is the mass-squared ratio. The loop functions are
\begin{equation}
  f_1(x)=\frac23x\left[(1+x)\ln\left(\frac{1+x}{x}\right)
        -\frac{2-x}{1-x}\right],
  \qquad
  f_2(x)=\frac{2}{3(x-1)} .
  \label{eq:f12}
\end{equation}

For the second term, which describes the washout from inverse decays, the washout function is
\begin{equation}
  W_i=\frac{K_i}{4}\left(\frac{M_i}{M_1}\right)z_i^3K_1(z_i).
  \label{eq:Wi}
\end{equation}
The flavor projection matrix $P^{(i)}$, determined by the flavor composition of the lepton state produced in the decay of $N_i$, is defined by
\begin{equation}
     c_{i\alpha}=\frac{h_{\alpha i}}{\sqrt{(h^\dagger h)_{ii}}},\qquad
  P^{(i)}_{\alpha\beta}=c_{i\alpha}c^\ast_{i\beta},
  \label{eq:ci}
\end{equation}
where $h_{\alpha i}$ is the coupling between the $i$-th heavy neutrino and the lepton doublet of flavor $\alpha$.

After obtaining the lepton asymmetry density matrix, the total $B-L$ asymmetry is computed by summing its diagonal elements, from which the baryon-to-photon ratio follows:
\begin{equation}
    \eta_B
    =\frac{28}{79}\frac{1}{27}
    \sum_\alpha N_{\alpha\alpha}
    \simeq 0.013
    \sum_\alpha N_{\alpha\alpha}.
    \label{eq:etaB}
\end{equation}
Here $28/79$ is the sphaleron conversion factor that converts the $B-L$ asymmetry into a baryon asymmetry in the SM~\cite{Khlebnikov1988,HarveyTurner1990}, and the factor $1/27$ is the normalization convention adopted for the equilibrium abundance $N_{N_i}^{\rm eq}=\frac{3}{8}z_i^2K_2(z_i)$. In the numerical calculation, we solve Eqs.~\eqref{eq:NN_Boltzmann} and~\eqref{eq:dmb3} simultaneously, including the contributions of the three heavy neutrinos, with the initial conditions
\begin{equation}
    N_{N_i}(z_{\rm in})=0,
    \qquad
    N_{\alpha\beta}(z_{\rm in})=0,
\end{equation}
which correspond to the assumption that the heavy-neutrino abundances and the lepton asymmetries vanish at the initial time. Since the density matrix $N_{\alpha\beta}$ is Hermitian, it contains three real diagonal elements and three complex off-diagonal elements, yielding nine independent real degrees of freedom. Solving the coupled equations then gives the evolution of the lepton asymmetry with the cosmic temperature and, ultimately, the baryon-to-photon ratio predicted by the model.

\subsection{Numerical results}
\label{subsec:numres}

\begin{figure}[htbp]
  \centering
  \includegraphics[width=\textwidth]{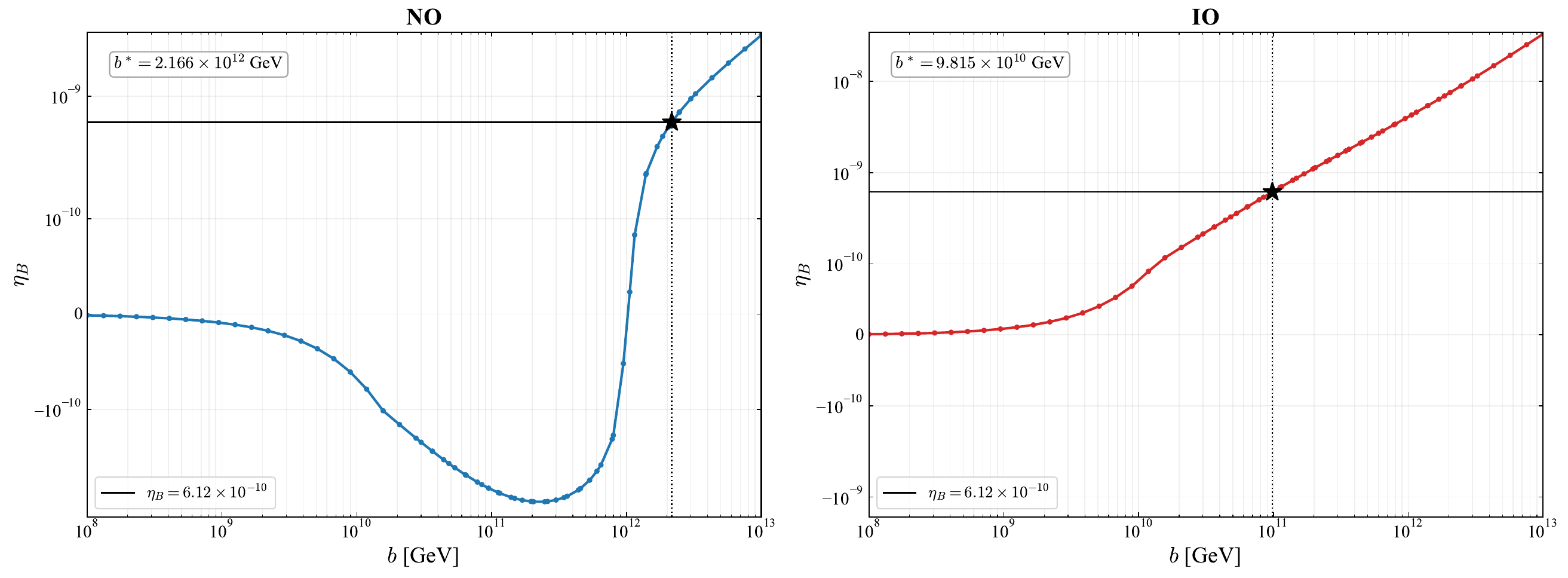} 
  \caption{The final baryon asymmetry $\eta_B$ as a function of the heavy-neutrino mass scale factor $b$, evaluated at the best-fit points for fermion masses and mixing in the NO (left) and IO (right) cases. Black stars mark the intersections of the curves with the observed value $\eta_B=6.12\times10^{-10}$~\cite{ParticleDataGroup:2024cfk}, and the corresponding scale factors $b^*$ that reproduce this value are indicated.}
  \label{fig:eta_B_vs_b}
\end{figure}

As discussed above, for a given benchmark point the scale factor $b$ of the right-handed neutrino mass matrix $M_N$ is the only free parameter, and its value uniquely determines the baryon asymmetry $\eta_B$. We adopt the best-fit points in Table~\ref{tab:bestfit_compare} as benchmark points and study leptogenesis for each of them. Varying $b$ and solving the Boltzmann equations~\eqref{eq:dmb3} and~\eqref{eq:NN_Boltzmann} gives the frozen value of $\eta_B$. The resulting curves of $\eta_B$ as a function of the heavy-neutrino scale $b$ are shown in Fig.~\ref{fig:eta_B_vs_b}.

In the NO case (left panel of Fig.~\ref{fig:eta_B_vs_b}), $\eta_B$ remains negative and decreases monotonically as $b$ increases. After reaching a minimum, it rises, crosses zero, and becomes positive. The observed asymmetry is reproduced at $b^*=2.166\times10^{12}$ GeV, corresponding to heavy-neutrino masses of order $10^{13}$ GeV, which are listed in Table~\ref{tab:heavy_masses}. In the IO case (right panel), by contrast, $\eta_B$ remains positive and increases monotonically with $b$, and the observed value is obtained at $b^*=9.815\times10^{10}$ GeV. The corresponding heavy-neutrino masses are about two orders of magnitude smaller than in the NO case, lying around $10^{11}$ GeV.
\begin{table}[htbp]
  \centering
  \begin{tabular}{|c|c|c|c|c|c|}
    \hline 
    Mass ordering& $b^*$ [GeV] & $a^*$ [GeV] & $M_1$ [GeV] & $M_2$ [GeV] & $M_3$ [GeV] \\
   \hline 
    NO & $2.166\times10^{12}$ & $23.09$ & $3.008\times10^{12}$ & $9.763\times10^{12}$ & $1.138\times10^{13}$ \\
    \hline 
    IO & $9.815\times10^{10}$ & $4.791$  & $1.345\times10^{11}$ & $1.503\times10^{11}$ & $1.678\times10^{11}$ \\
    \hline 
  \end{tabular}
    \caption{ Heavy-neutrino masses and the corresponding scale factors $b^*$ and $a^*$ at the points that reproduce the observed baryon asymmetry $\eta_B=6.12\times10^{-10}$.}
  \label{tab:heavy_masses}
\end{table}

\begin{figure}[ht]
  \centering
  \includegraphics[width=\textwidth]{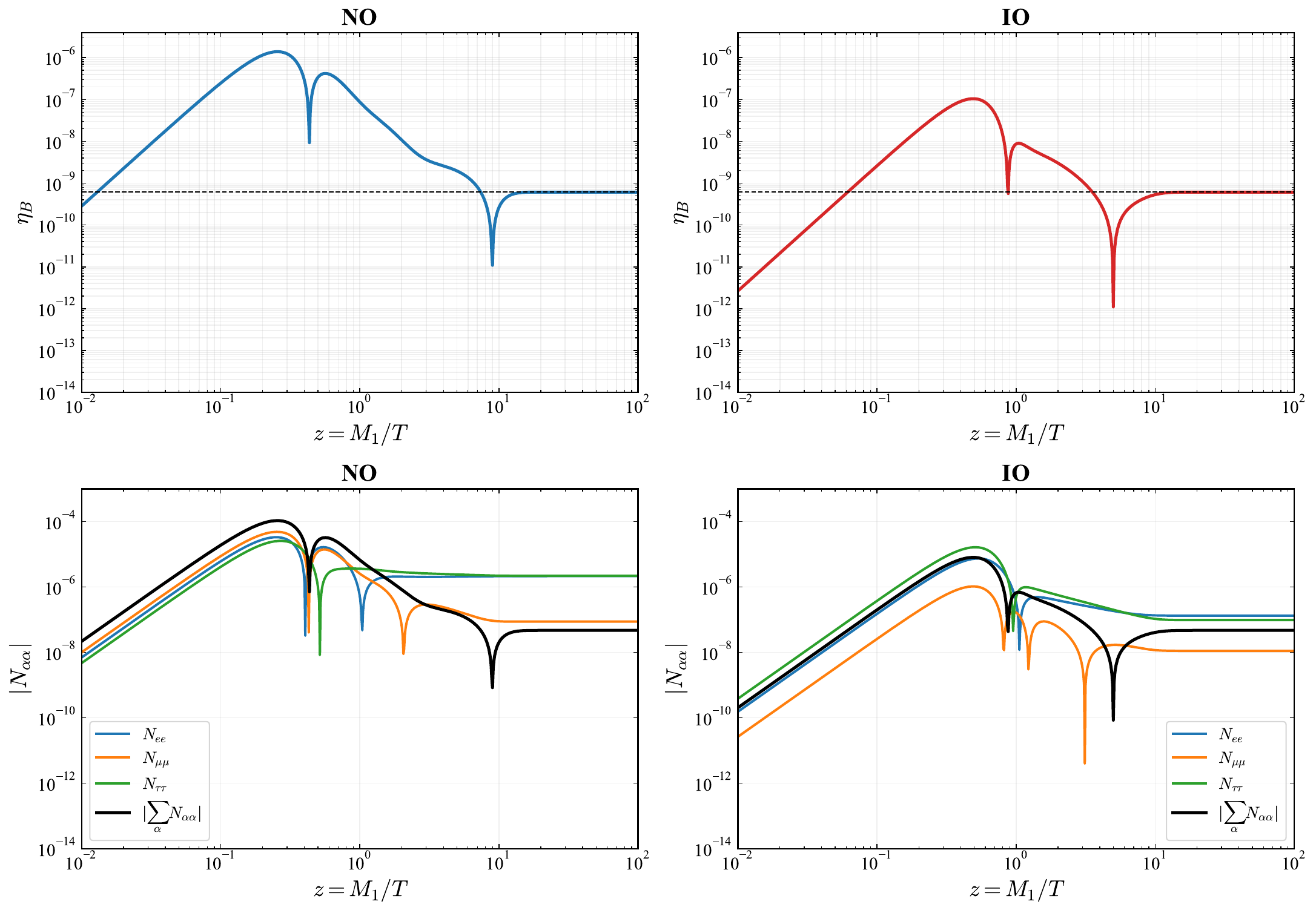}
  \caption{ Evolution of the lepton asymmetry as a function of $z\equiv M_1/T$. The parameters are chosen so that the final values match the observed baryon asymmetry $\eta_B=6.12\times10^{-10}$ (NO: $b^*=2.166\times10^{12}$ GeV; IO: $b^*=9.815\times10^{10}$ GeV). Top row: $|\eta_B(z)|$ versus $z$; the absolute values are used to display the order of magnitude, and the dotted line indicates the observed value. Bottom row: the asymmetries of the three individual flavors, $|N_{ee}|$, $|N_{\mu\mu}|$, and $|N_{\tau\tau}|$, together with their sum $|\sum_\alpha N_{\alpha\alpha}|$ (black line). The left and right columns correspond to the NO and IO cases, respectively.}
  \label{fig:evolution}
\end{figure}

For the NO and IO cases, we select the scale $b^*$ that reproduces the observed value of $\eta_B$ and solve for the evolution of $\eta_B$ with respect to $z$, as shown in Fig.~\ref{fig:evolution}. The figure illustrates the complete evolution of the lepton asymmetry from the very early Universe ($T\gg M_1$) down to low temperatures ($T\ll M_1$), evaluated at a parameter point that simultaneously satisfies the neutrino oscillation data and the observed baryon asymmetry. In the upper panels, at $z\lesssim 0.1$, the heavy neutrinos have not yet been abundantly produced ($N_i\ll N_i^{\rm eq}$). As the heavy neutrinos fall out of thermal equilibrium, the lepton asymmetry is rapidly established and reaches a peak around $z\simeq 0.25$ (NO) and $z\simeq 0.5$ (IO). The washout from inverse decays then gradually dominates. Finally, as the temperature decreases, the asymmetry freezes out at the observed value $\eta_B=6.12\times10^{-10}$.

The lower panels of Fig.~\ref{fig:evolution} show the contributions of the three individual flavors. The asymmetry magnitudes of a single flavor (e.g., $e$ and $\tau$) are of order $10^{-6}$ (NO) and $10^{-7}$ (IO), whereas their sum is only of order $10^{-8}$. This indicates that the final baryon asymmetry is a residual remaining after significant cancellation among the different flavors, which is why solving the complete density matrix equations is essential to calculate it reliably.

\begin{figure}[ht]
  \centering
  \includegraphics[width=\textwidth]{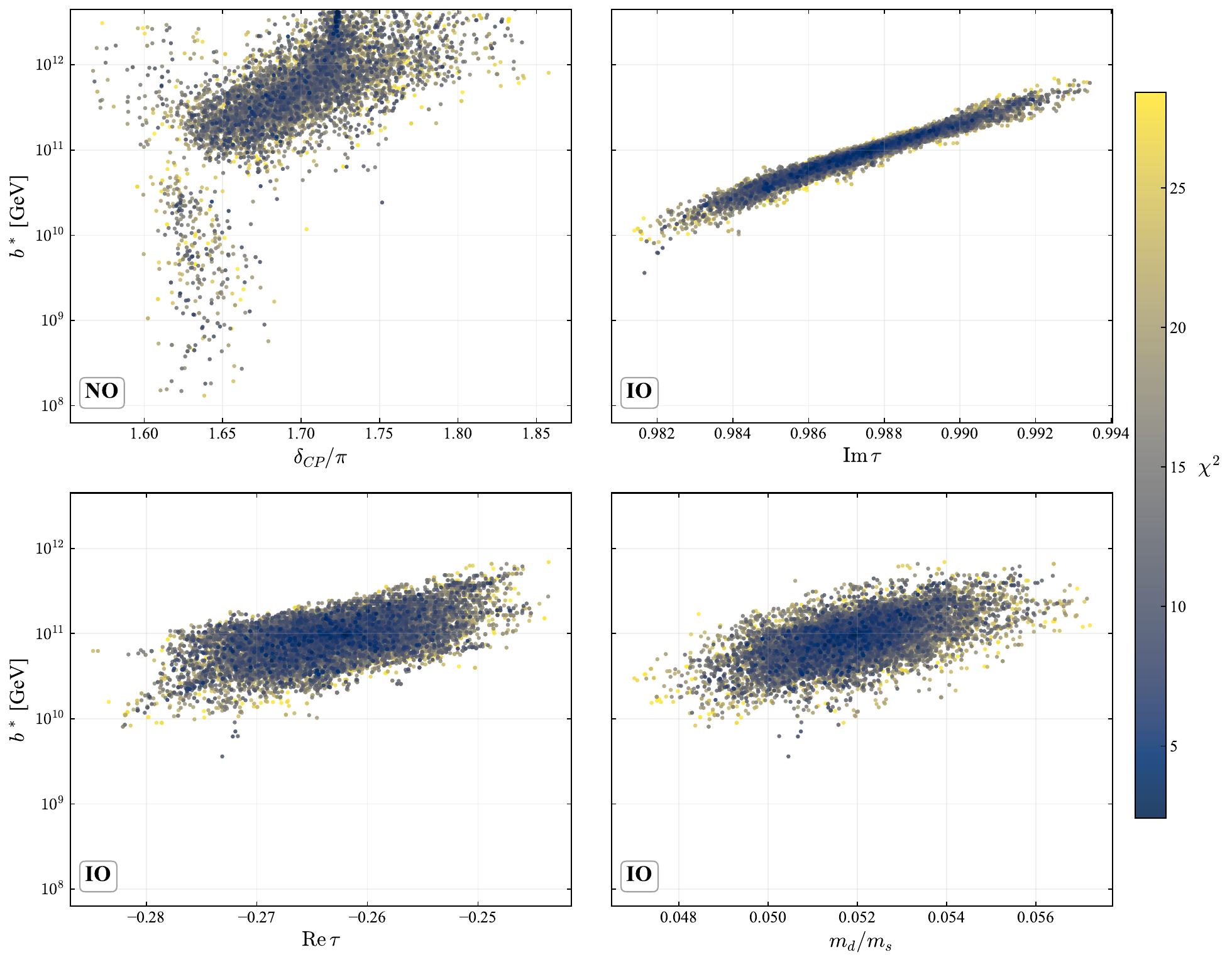}
  \caption{Correlations between the scale factor $b^*$, which reproduces the observed baryon asymmetry, and the model parameters. All data points shown satisfy $\chi^2<30$, and the equation $\eta_B(b)=6.12\times10^{-10}$ has real solutions. The panels show the correlations between $b^*$ and leptonic $\delta_{\mathrm{CP}}$ in the NO case and between $b^*$ and $\tau$ and $m_d/m_s$ in the IO case. The color of the scan points indicates the $\chi^2$ value.}
  \label{fig:bstar}
\end{figure}

Having analyzed the viability of leptogenesis at the best-fit points, we now investigate all datasets obtained with the MCMC method that satisfy $\chi^2<30$, as points with excessively large $\chi^2$ are in significant tension with the experimental data. For each dataset, since $b$ is the only free parameter, the final baryon asymmetry can be treated as a function of $b$, $\eta_B\equiv\eta_B(b)$. Solving the Boltzmann equations numerically, we denote by $b=b^*$ any real solution of $\eta_B(b)=6.12\times10^{-10}$. This procedure identifies all parameter points that account for the observed baryon asymmetry. Figure~\ref{fig:bstar} displays the results, showing the cases where $b^*$ exhibits significant correlations with other parameters.

In the NO case, $b^*$ shows a correlation with the leptonic Dirac CP-violating phase $\delta_{\mathrm{CP}}$. For $1.6\pi<\delta_{\mathrm{CP}}<1.7\pi$, $b^*$ spans a broad range from $10^8$ to $10^{12}$ GeV, whereas for $\delta_{\mathrm{CP}}<1.6\pi$ or $\delta_{\mathrm{CP}}>1.7\pi$, $b^*$ is generally larger, typically above $10^{10}$ GeV. This suggests that, if future experiments determine the leptonic Dirac CP phase precisely, the model could provide specific predictions for the heavy-neutrino mass scale.

In the IO case, $b^*$ shows a correlation with the modulus VEV $\langle\tau\rangle$, being positively correlated with both $\mathrm{Re}\,\tau$ and $\mathrm{Im}\,\tau$. This may indicate that modular symmetry connects the fermion mass mixing to the cosmic baryon asymmetry. The breaking of the modulus VEV can generate the hierarchical structure of the heavy neutrinos while simultaneously inducing their CP asymmetries. In addition, $b^*$ shows a weaker positive correlation with the quark mass ratio $m_d/m_s$. Since the experimental measurements of the quark mass ratios are relatively precise, this stringent constraint may result in a narrower allowed range for $b^*$ in the IO case than in the NO case.

\section{Summary and outlook}
\label{sec:conclusions}
In this paper, we have constructed a non-SUSY $SU(5)$ grand unified model based on the non-holomorphic $S'_4$ modular flavor symmetry and a global $U(1)_{PQ}$ symmetry. The model aims to explain the SM fermion masses and mixing as well as the observed baryon asymmetry of the Universe within a unified framework.

By introducing the non-holomorphic $S'_4$ modular symmetry, the fermion Yukawa couplings are restricted to polyharmonic Maa{\ss} forms, which are functions of the modulus $\tau$. When the modulus acquires a VEV, the modular symmetry is broken, thereby determining the structure of the Yukawa couplings. Based on current experimental data, we performed a systematic numerical analysis of the fermion masses and mixing. The results show that the model provides excellent fits in both the NO and IO cases, with all Pull values satisfying $|\mathcal{P}| < 0.8$.

The model further yields specific predictions for yet-unmeasured parameters. In the NO case, the predicted absolute neutrino masses are $m_1 \simeq 1.91$ meV, $m_2 \simeq 8.89$ meV, and $m_3 \simeq 50.1$ meV, with the Dirac CP phase $\delta_{\mathrm{CP}} \simeq 1.72\pi$. For the Majorana phases, $\eta_1$ exhibits a clear bimodal structure, whereas $\eta_2$ is predominantly single-peaked. The predicted effective Majorana mass is extremely small ($m_{\beta\beta} \simeq 2.42$ meV), well below the reach of current and next-generation $0\nu\beta\beta$ decay experiments. In the IO case, by contrast, the mass spectrum is highly degenerate ($m_1 \simeq 49.1$ meV, $m_2 \simeq 49.9$ meV, and $m_3 \simeq 2.19$ meV), and $\delta_{\mathrm{CP}} \simeq 0.77\pi$. Here $\eta_2$ becomes bimodal while $\eta_1$ is single-peaked. The predicted value $m_{\beta\beta} \simeq 24.6$ meV lies below the current most stringent limit from KamLAND-Zen but falls right at the edge of the target sensitivity windows of next-generation $0\nu\beta\beta$ experiments such as nEXO and LEGEND-1000.

We also investigated thermal leptogenesis using the complete density matrix equations. Once the fitted parameters for fermion masses and mixing are fixed, the heavy-neutrino mass scale is determined by the single free parameter $b$. We find that, for both the NO and IO cases, a specific value of $b$ reproduces the observed baryon asymmetry $\eta_B = 6.12\times10^{-10}$, indicating that $b$ provides the link between the heavy-neutrino mass scale and the cosmic baryon asymmetry. The numerical results show that at the best-fit points the NO case requires heavy-neutrino masses of $\mathcal{O}(10^{13})$ GeV, while the IO case requires $\mathcal{O}(10^{11})$ GeV. The flavor-space evolution further reveals significant cancellations among the different lepton flavors, demonstrating the need for the full density matrix treatment. We also find strong correlations between the scale factor $b^*$ and quantities such as the Dirac CP phase and the modulus VEV, providing a potential bridge between low-energy flavor physics and early-Universe CP violation.

As phenomenological complements to this unified framework, we briefly discussed GCU, proton decay, and axion physics in Appendix~\ref{sec:gutpq}. Because of the selection rules of the $U(1)_{PQ}$ symmetry, the model avoids dangerous proton decay operators. In addition, the breaking of the $U(1)_{PQ}$ symmetry at the GUT scale naturally yields an ultralight axion that is a viable dark matter candidate.

Although the model is predictive, several avenues remain for future exploration. Future neutrino experiments such as DUNE~\cite{DUNE:2020ypp}, Hyper-Kamiokande~\cite{Hyper-Kamiokande:2018ofw}, and JUNO~\cite{JUNO:2024jaw} will provide more precise measurements of the leptonic CP phase $\delta_{\mathrm{CP}}$ and determine the neutrino mass ordering, providing stringent tests of the parameter space and of the $b^*$ predictions obtained in this paper. Future $0\nu\beta\beta$ decay experiments will probe the predicted $m_{\beta\beta}$ range and further distinguish between the NO and IO cases. Moreover, although the high-energy states predicted here lie in a mass region that is difficult to access with conventional laboratory searches, the model reveals strong correlations between $b^*$ and low-energy observables such as the Dirac CP phase, which provide a potential handle for future experimental tests. Finally, our leptogenesis analysis has not yet included effects such as $\Delta L=1$ scattering terms and finite-temperature corrections. A more detailed investigation of these effects would further refine the quantitative predictions for the baryon asymmetry.

\section*{Acknowledgments}
This work was supported by the National Natural Science Foundation of China under Grant No.~12565016.


\appendix

\section{Phenomenological implications of the GUT and the PQ symmetry}
\label{sec:gutpq}
Although the model reproduces the observed fermion masses and mixing and the baryon asymmetry of the Universe, a complete assessment of a grand unified model also requires examining its predictions for gauge coupling unification and proton decay. In addition, the breaking of the $U(1)_{PQ}$ symmetry generates an ultralight pseudo-Nambu--Goldstone boson, the axion, which may address two major shortcomings of the SM: the strong CP problem and the nature of dark matter. We therefore discuss how the model achieves gauge coupling unification, how it satisfies the constraints from proton decay, and what axion mass it predicts.

\subsection{Gauge coupling unification}
\label{subsec:gcu}
The Higgs sector of the present  $SU(5)\times S'_4\times U(1)_{PQ}$ model contains $H_5$, $H_{45}$, $H'_5$ and $H_{24}$. The electroweak doublets contained in these $SU(5)$ multiplets are
\begin{equation}
H_1\subset H_5,\qquad H_2\subset H_{45},\qquad H_3\subset H'_{5},
\end{equation}
with $H_1,H_2,H_3\sim(1,2,+1/2)$. These three doublets can mix through the scalar potential, so we diagonalize the corresponding doublet mass matrix as
\begin{equation}
U_H^\dagger\,\mathcal{M}_H^2\,U_H =
\operatorname{diag}
\left(m_h^2,\,M_{H_A}^2,\,M_{H_B}^2 \right),
\end{equation}
where $h$ denotes the SM Higgs boson with $m_h\simeq125$ GeV, and the two additional physical doublets $H_A$ and $H_B$ contribute threshold corrections to the gauge coupling running. The states contained in $45_H$ are denoted by
\begin{equation}
\phi_3\sim(3,3,-1/3),
\qquad
\phi_5\sim(8,2,+1/2).
\end{equation} 
The scalar states relevant for the unification analysis are summarized in Table~\ref{tab:GCU_fields}.
\begin{table}[htbp]
    \centering
    \renewcommand{\arraystretch}{1.6}
    \begin{tabular}{|c|c|c|cc|}
    \hline
        States & $SU(3)_C\times SU(2)_L\times U(1)_Y$
        & $\Delta b_1$ & $\Delta b_2$ & $\Delta b_3$ \\
        \hline
        $h$  &$(1,2,+1/2)$ &\multicolumn{3}{c|}{included in $b_i^{\rm SM}$}\\
    \hline
        $H_A$  &$(1,2,+1/2)$ &$1/10$ & $1/6$ & $0$\\
        $H_B$  &$(1,2,+1/2)$ &$1/10$ & $1/6$ & $0$\\
        $\phi_3$ &$(3,3,-1/3)$ &$1/5$ & $2$ & $1/2$\\
        $\phi_5$ &$(8,2,+1/2)$ &$4/5$ & $4/3$ & $2$\\
    \hline
    \end{tabular}
    \caption{Scalar states relevant to gauge coupling unification. Here $H_A$ and $H_B$ denote the two additional physical electroweak-doublet mass eigenstates, and $\Delta b_i$ are their contributions to the one-loop beta-function coefficients. The remaining scalar components are assumed to lie at or near the GUT scale $M_X$ and are omitted, since their threshold corrections are negligible.}
    \label{tab:GCU_fields}
\end{table}

We study gauge coupling unification at one-loop order using the standard formalism~\cite{GiveonHallSarid1991}. At this order, the gauge couplings obey
\begin{equation}
\frac{d\,\alpha_i^{-1}}{d\ln\mu}=-\frac{b_i}{2\pi},
\qquad b_i(\mu)=b_i^{\rm SM}+\sum_{M_I<\mu}\Delta b_i^I ,\end{equation}
where $b_i^{\rm SM}=\left(\frac{41}{10},-\frac{19}{6},-7\right)$ are the SM beta-function coefficients and $I$ runs over the additional scalar states whose masses $M_I$ lie below the scale $\mu$. Integrating from $M_Z$ to the GUT scale $M_X$ gives
\begin{equation}
\alpha_i^{-1}(M_Z)=\alpha_X^{-1}+\frac{b_i^{\rm SM}}{2\pi} \ln\frac{M_{\rm GUT}}{M_Z}+\frac{1}{2\pi}\sum_I\Delta b_i^I \ln\frac{M_X}{M_I}.
\label{eq:GCU_master}
\end{equation}

The SM Higgs $h$ is already included in $b_i^{\rm SM}$, so $H_A$, $H_B$, $\phi_3$, and $\phi_5$ enter as additional sub-GUT thresholds. The remaining scalar components and the heavy $SU(5)$ gauge bosons $X$ and $Y$ are assumed to lie at the unification scale $M_X$. At this scale, the three gauge couplings are expected to be equal, $\alpha_X=\alpha_1=\alpha_2=\alpha_3$. Using this condition, for a given $M_X$ we can invert Eq.~\eqref{eq:GCU_master} to extract the masses of the scalar fields listed in Table~\ref{tab:GCU_fields} as well as the value of $\alpha_X$.

We first define
\begin{equation}
x_I\equiv \ln\frac{M_X}{M_I},\qquad L\equiv\ln\frac{M_X}{M_Z}.
\end{equation}
Taking differences of Eq.~\eqref{eq:GCU_master} then yields two independent unification conditions,
\begin{align}
D_{12}&=-\frac95x_{\phi_3}-\frac{8}{15}x_{\phi_5}-\frac1{15}(x_{H_A}+x_{H_B}),
\label{eq:GCU12_compact}\\
D_{23}&=\frac32x_{\phi_3}-\frac23x_{\phi_5}+\frac16(x_{H_A}+x_{H_B}),
\label{eq:GCU23_compact}
\end{align}
where
\begin{equation}
D_{12}=2\pi\left[\alpha_1^{-1}(M_Z)-\alpha_2^{-1}(M_Z)\right]-\left(b_1^{\rm SM}-b_2^{\rm SM}\right)L,
\end{equation}
and
\begin{equation}
D_{23}=2\pi\left[\alpha_2^{-1}(M_Z)-\alpha_3^{-1}(M_Z)\right]-\left(b_2^{\rm SM}-b_3^{\rm SM}\right)L.
\end{equation}

For given values of $M_{\rm GUT}$, $M_{H_A}$, and $M_{H_B}$, Eqs.~\eqref{eq:GCU12_compact} and~\eqref{eq:GCU23_compact} determine the colored threshold masses $M_{\phi_3}$ and $M_{\phi_5}$. Defining
\begin{equation}
R_{12}=D_{12}+\frac1{15}(x_{H_A}+x_{H_B}),\qquad R_{23}=D_{23}-\frac16(x_{H_A}+x_{H_B}),
\end{equation}
we obtain
\begin{equation}
x_{\phi_3}=-\frac{1}{3}R_{12}+\frac{4}{15}R_{23},\qquad x_{\phi_5}=-\frac{3}{4}R_{12}-\frac{9}{10}R_{23},
\end{equation}
and hence
\begin{equation}
M_{\phi_3}=M_Xe^{-x_{\phi_3}},\qquad M_{\phi_5}=M_Xe^{-x_{\phi_5}}.
\end{equation}
The unified gauge coupling is determined from the $SU(3)_C$ coupling,
\begin{equation}
\alpha_X^{-1}=\alpha_3^{-1}(M_Z)-\frac{b_3^{\rm SM}}{2\pi}L-\frac{1}{2\pi}\sum_I\Delta b_3^I x_I,
\end{equation}
which in the present case reduces to
\begin{equation}
\alpha_X^{-1}=\alpha_3^{-1}(M_Z)+\frac{7}{2\pi}L-\frac{1}{2\pi}\left(\frac12x_{\phi_3}+2x_{\phi_5}\right).
\label{eq:alphaGUT_compact}
\end{equation}

\begin{figure}[htbp]
    \centering
    \includegraphics[width=\textwidth]{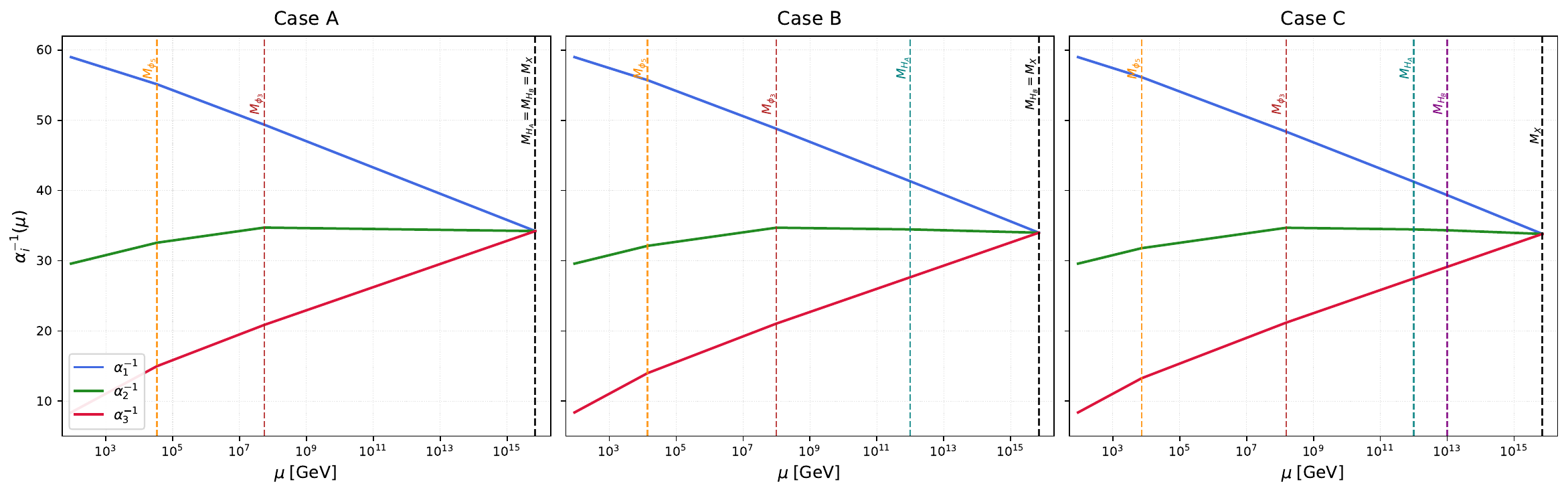}
    \caption{One-loop gauge coupling evolution for the three representative scalar-threshold realizations. The panels correspond to Cases A, B, and C, respectively. The vertical dashed lines indicate the corresponding scalar thresholds and the GUT scale.}
    \label{fig:GCU_three_cases}
\end{figure}

We fix $M_X=7.0\times10^{15}$ GeV and consider three representative realizations of the physical scalar-threshold spectrum, as summarized in Table~\ref{tab:GCU_benchmarks}. These cases correspond to (A) both $H_A$ and $H_B$ at the GUT scale, (B) one of them at an intermediate scale, and (C) both below $M_X$. The three cases yield slightly different values of the unified coupling $\alpha_X$, although these values remain relatively close, especially for Cases B and C. The corresponding gauge coupling evolution curves are shown in Fig.~\ref{fig:GCU_three_cases}.
\begin{table}[htbp]
\centering
\begin{tabular}{|c|ccccc|}
\hline
Case & $M_{H_A}$ & $M_{H_B}$ & $M_{\phi_3}$ & $M_{\phi_5}$ & $\alpha_X^{-1}$ \\
\hline
A & $M_X$ & $M_X$ & $5.469\times10^{7}$ & $3.370\times10^{4}$ & $34.220959$ \\
B & $1.000\times10^{12}$ & $M_X$ & $9.869\times10^{7}$ & $1.390\times10^{4}$ & $33.986109$ \\
C & $1.000\times10^{12}$ & $1.000\times10^{13}$ & $1.527\times10^{8}$ & $7.220\times10^{3}$ & $33.812336$ \\
\hline
\end{tabular}
\caption{Representative scalar-threshold spectra that satisfy one-loop gauge coupling unification for $M_X=7.0\times10^{15}$ GeV. All masses are given in GeV. The states $H_A$ and $H_B$ are the two additional physical electroweak-doublet mass eigenstates detailed in Table~\ref{tab:GCU_fields}.}
\label{tab:GCU_benchmarks}
\end{table}

\subsection{Axion mass and the Peccei--Quinn scale}
\label{subsec:axpq}

The vacuum expectation value of the complex scalar $24_H$,
\begin{equation}
\langle 24_H\rangle=\frac{v_\Sigma}{\sqrt{2}}\frac{1}{\sqrt{30}}\mathrm{diag}(2,2,2,-3,-3),
\end{equation}
simultaneously breaks the $SU(5)$ and $U(1)_{PQ}$ symmetries, linking the Peccei--Quinn scale to the GUT scale~\cite{FileviezPerez:2019ssf}. With the normalization adopted here, the axion decay constant $f_a$ is
\begin{equation}
f_a\simeq\frac{v_\Sigma}{6}.
\end{equation}
The heavy gauge bosons acquire the mass
\begin{equation}
M_X= M_{\rm GUT}=\sqrt{\frac{10\pi}{3}\alpha_{\rm GUT}}\,v_\Sigma,
\end{equation}
which yields the relation
\begin{equation}
f_a=\frac{M_X}{\sqrt{\alpha_{\rm GUT}}}\frac{1}{2\sqrt{30\pi}}.
\label{eq:fa_MX}
\end{equation}
Thus, once $M_X$ and $\alpha_{\rm GUT}$ are fixed by gauge coupling unification, $f_a$ is uniquely determined.

We adopt Benchmark Point B, which corresponds to
\begin{equation}
M_X=7.0\times10^{15}\ {\rm GeV},\qquad\alpha_{\rm GUT}^{-1}=33.986109,
\end{equation}
and yields
\begin{equation}
f_a\simeq2.10\times10^{15}\ {\rm GeV}.
\label{eq:fa_B}
\end{equation}
Using the standard QCD axion mass–decay-constant relation~\cite{GorghettoVilladoro2019},
\begin{equation}
m_a=5.691(51)\times10^{-6}\ {\rm eV}\left(\frac{10^{12}\ {\rm GeV}}{f_a}\right),
\label{eq:axion_mass}
\end{equation}
we obtain
\begin{equation}
m_a\simeq2.71\times10^{-9}\ {\rm eV}.
\end{equation}

Next, we consider the pre-inflationary PQ-breaking scenario, in which the observable Universe is characterized by a single initial misalignment angle $\theta_i$, corresponding to the initial field value $a_i=f_a\theta_i$. Assuming standard radiation-dominated cosmology, the axion abundance produced through the misalignment mechanism is approximately~\cite{VisinelliGondolo2009}
\begin{equation}
\Omega_a h^2
\simeq
0.12\,\theta_i^2 F(\theta_i)
\left(
\frac{f_a}{5\times10^{11}\ {\rm GeV}}
\right)^{7/6},
\label{eq:axion_abundance}
\end{equation}
where
\begin{equation}
F(\theta_i) =\left[\ln\left(\frac{e}{1-\theta_i^2/\pi^2}\right)\right]^{7/6}.
\end{equation}
Requiring the axion to account for the observed dark matter abundance, $\Omega_a h^2\simeq0.12$~\cite{Planck:2018vyg}, then yields the initial misalignment angle $\theta_i \simeq 7.70\times10^{-3}$ for $f_a\simeq2.10\times10^{15}$ GeV.

\subsection{Proton decay}
\label{subsec:protdcy}
Two types of dimension-six proton decay operators can arise in our model, induced by colored-scalar exchange and by heavy gauge boson exchange. The relevant colored state is $\phi_3\sim(3,3,-1/3)$, contained in $45_H$. In a generic $SU(5)$ model, $\phi_3$-mediated proton decay would require the couplings
\begin{equation}
10_F\,\overline{5}_F\,45_H^\ast,\quad10_F\,10_F\,45_H.
\end{equation}
In our model, however, the second operator is forbidden by the $U(1)_{PQ}$ symmetry, since $Q_{PQ}(10_F10_F45_H)\neq0$, the $\phi_3$-mediated contribution is absent at the renormalizable level. Possible higher-dimensional contributions are not considered here. The leading contribution is therefore mediated by the heavy $SU(5)$ gauge bosons $X$ and $Y$.

Integrating out $X$ and $Y$ gives the effective Lagrangian
\begin{equation}
\mathcal{L}_{\rm eff}^{(X,Y)}\sim
\frac{g_{\rm GUT}^2}{M_X^2}(qq)(q\ell),
\end{equation}
so that, up to hadronic and flavor factors, $\tau_p\propto M_X^4/\alpha_{\rm GUT}^2m^5_p$. This is the standard dimension-six gauge-boson contribution in $SU(5)$~\cite{FileviezPerez2004,FileviezPerez:2019ssf}. The Super-Kamiokande Collaboration reports the partial-lifetime bound $\tau_p(p\to e^+\pi^0)>2.4\times10^{34}$ yr at $90\%$ C.L.~\cite{SuperKamiokande2020}, which, using the proton-decay analysis adopted here, corresponds to $M_X\gtrsim5\times10^{15}$ GeV. Our benchmark value $M_X=7.0\times10^{15}$ GeV therefore satisfies the proton-decay constraint.

\section{\texorpdfstring{$S'_4$}{S4prime} Modular symmetry}
\label{app:modular}
In this appendix, we briefly review the basic concepts of $S'_4$ modular symmetry and the properties of non-holomorphic modular forms. These ingredients are essential for constructing the modular-invariant Yukawa couplings and the fermion mass matrices in the main text.

The modular group is $\Gamma=SL(2,\mathbb{Z})$, which acts on the complex modulus $\tau$, with $\operatorname{Im}\tau>0$, through linear fractional transformations,
\begin{equation}
\gamma:\tau\rightarrow\gamma\tau=\frac{a\tau+b}{c\tau+d},\qquad\gamma=
\begin{pmatrix}
a&b\\c&d
\end{pmatrix}
\in\Gamma.
\end{equation}
The modular group is generated by
\begin{equation}
S=\begin{pmatrix}0&1\\-1&0\end{pmatrix},
\qquad
T=\begin{pmatrix}1&1\\0&1\end{pmatrix},
\end{equation}
which satisfy $S^4=(ST)^3=\mathbf{I}$ and $S^2T=TS^2$, where $\mathbf{I}$ is the $2\times2$ identity matrix.

The principal congruence subgroup of level $N$ is defined by
\begin{equation}
\Gamma(N)=\left\{\gamma\in\Gamma\ \big|\ \gamma\equiv \mathbf{I}\pmod N \right\},
\end{equation}
with $N\in\mathbb{Z}^+$. The finite modular group is $\Gamma_N=\Gamma/\Gamma(N)$. For $N>2$, its double covering group is denoted by $\Gamma_N'=\Gamma/\Gamma(N)$. For $N=4$, as employed in this work, one has
\begin{equation}
\Gamma_4\simeq S_4,\quad \Gamma_4'\simeq S_4',
\end{equation}
where $S'_4$ is the binary octahedral group. Its generators satisfy $S^2=R$, $(ST)^3=T^4=R^2=\mathbf{I}$, $RT=TR$, with $R=S^2$.

 The group $S'_4$ has ten irreducible representations: $\mathbf{1}$, $\mathbf{1}'$, $\hat{\mathbf{1}}$, $\hat{\mathbf{1}}'$, $\mathbf{2}$, $\hat{\mathbf{2}}$, $\mathbf{3}$, $\mathbf{3}'$, $\hat{\mathbf{3}}$, and $\hat{\mathbf{3}}'$. The hatted representations are unique to the double cover $S'_4$ and have no counterpart in $S_4$, whereas the unhatted representations are common to both groups. The multiplication rules for the irreducible representations relevant to this work are
\begin{equation}
    \begin{aligned}
        & \mathbf{\hat{1}} \otimes \mathbf{\hat{1}}' =\mathbf{1} \otimes \mathbf{1} =\mathbf{1}, \quad \mathbf{\hat{1}} \otimes \mathbf{1}'=\mathbf{\hat{1}}'\quad \mathbf{1}' \otimes \mathbf{3} = \mathbf{3}', \quad \mathbf{1}' \otimes \mathbf{3}' = \mathbf{3}, \\
        &  \hat{\mathbf{1}} \otimes \mathbf{3} =\hat{\mathbf{1}}' \otimes \mathbf{3}'= \hat{\mathbf{3}}, \quad \hat{\mathbf{1}}' \otimes \mathbf{3} =\hat{\mathbf{1}} \otimes \mathbf{3}'= \hat{\mathbf{3}'},\quad \mathbf{2} \otimes \mathbf{2} = \mathbf{1} \oplus \mathbf{1}'\oplus\mathbf{2}, \\
        & \mathbf{3} \otimes \mathbf{3} =\hat{\mathbf{3}} \otimes \hat{\mathbf{3}}'= \mathbf{1} \oplus \mathbf{2} \oplus \mathbf{3} \oplus \mathbf{3}',  \quad \mathbf{3} \otimes \mathbf{3}' = \mathbf{1}' \oplus \mathbf{2} \oplus \mathbf{3} \oplus \mathbf{3}'.
    \end{aligned}
\end{equation}
A complete list of multiplication rules and the associated Clebsch--Gordan coefficients can be found in Ref.~\cite{Ding:2023htn}.

In modular flavor models, the modulus $\tau$ is a dynamical field. Its VEV breaks the modular symmetry and fixes the numerical values of the modular forms. Matter and Higgs fields transform under a modular transformation as
\begin{equation}
\psi_i\rightarrow
(c\tau+d)^{-k_i}\rho_i(\gamma)\psi_i,\qquad H\rightarrow(c\tau+d)^{-k_H}\rho_H(\gamma)H,
\end{equation}
where $k_i$ and $k_H$ are their modular weights and $\rho_i(\gamma)$ and $\rho_H(\gamma)$ are the corresponding representation matrices. For a Yukawa interaction involving matter fields $\psi_i$ and a Higgs field $H$, the associated modular form must compensate the modular weights of these fields. If the modular form has weight $k_Y$, modular invariance therefore requires $k_Y=k_H+\sum_i k_i$, which ensures that the complete Yukawa interaction is invariant.

In conventional supersymmetric modular flavor models, the Yukawa couplings are constructed from holomorphic modular forms. In non-SUSY models, the holomorphicity requirement can be relaxed, and non-holomorphic modular forms depending on both $\tau$ and $\bar\tau$ can be used. A non-holomorphic modular form of weight $k$ and level $N$ satisfies $Y(\gamma\tau)=(c\tau+d)^kY(\tau)$. In the framework considered here, these functions also satisfy the harmonic condition $\Delta_kY=0$, where the weight-$k$ Laplacian is~\cite{Ding:2023htn}
\begin{equation}
\Delta_k=-4y^2\partial_\tau\partial_{\bar\tau}
+2iky\partial_{\bar\tau},
\qquad \tau=x+iy.
\end{equation}
They additionally obey a moderate-growth condition at the cusp, $Y(x+iy)=O(y^\alpha)$ as $y\rightarrow\infty$, where $\alpha$ is a real constant. These properties characterize the non-holomorphic modular forms considered in this work. Such a form can be expanded in a Fourier series as~\cite{Qu:2025ddz}
\begin{equation}
Y(\tau)=\sum_{\substack{n\geq 0\\n\in\frac{1}{N}\mathbb Z}}
c^+(n)q^n+\sum_{\substack{n\leq 0\\n\in\frac{1}{N}\mathbb Z}}
c^-(n)\beta_{1-k}(n,y)q^n,
\end{equation}
where $q=e^{2\pi i\tau}$ and
\begin{equation}
\beta_{1-k}(n,y)=
\begin{cases}
\Gamma(1-k,-4\pi ny),&n\neq0,\\
y^{1-k},&n=0,\ k\neq1,\\
\log y,&n=0,\ k=1.
\end{cases}
\end{equation}
Here $\Gamma(s,x)$ is the incomplete gamma function. Non-holomorphic modular forms can also be arranged into multiplets transforming in irreducible representations of the finite modular group.

A modular multiplet $Y_{\mathbf r}^{(k)}(\tau)$ transforms under $\gamma\in\Gamma$ as
\begin{equation}
Y_{\mathbf r}^{(k)}(\gamma\tau) =(c\tau+d)^k \rho_{\mathbf r}(\gamma) Y_{\mathbf r}^{(k)}(\tau),
\end{equation}
where $\mathbf{r}$ labels an irreducible representation of the finite modular group and $\rho_{\mathbf{r}}(\gamma)$ is its representation matrix. For the level-four case, Refs.~\cite{Qu:2024rns,Qu:2025ddz} provide the explicit $q$-expansions of the non-holomorphic modular forms and show how these are organized into irreducible representations of $S'_4$. For weights $k>2$, the non-holomorphic modular forms coincide with the holomorphic modular forms. Their explicit $q$-expansions can be found in Ref.~\cite{Ding:2023htn}. For $k\le2$, the even-weight forms can be arranged into the representations $\mathbf{1}$, $\mathbf{2}$, and $\mathbf{3}$, whereas the odd-weight forms can be arranged into $\hat{\mathbf{3}}$ and $\hat{\mathbf{3}}'$. These representations provide the modular multiplets relevant to constructing the Yukawa structures of the model.

To construct an $S'_4$-invariant Yukawa interaction, the tensor product of the representations carried by the modular form, the matter fields, and the Higgs field must contain the trivial singlet, satisfying the condition $\mathbf r_Y\otimes\left(\bigotimes_i\mathbf r_{\psi_i}\right)\otimes\mathbf r_H\supset\mathbf{1}$. The invariant contraction of these multiplets determines the allowed Yukawa structures. After the modulus $\tau$ acquires a VEV, the modular multiplets become numerical functions, and the resulting Yukawa couplings determine the fermion mass matrices.

%
%

\bibliographystyle{RAN}
\bibliography{biblio}

\end{document}